\documentclass[fleqn,usenatbib,useAMS]{mnras}

\usepackage[utf8]{inputenc} 
\usepackage[T1]{fontenc}    
\usepackage{hyperref}       
\usepackage{url}            
\usepackage{booktabs}       
\usepackage{amsfonts}       
\usepackage{nicefrac}       
\usepackage{microtype}      
\usepackage{lineno}
\usepackage{natbib}
\usepackage{pifont}
\usepackage{amsmath,amssymb}
\usepackage{graphicx}
\usepackage{caption}
\usepackage{subcaption}
\usepackage{multicol,graphicx,xcolor}
\usepackage{tikz}
\usetikzlibrary{shapes.misc, positioning}
\usetikzlibrary{matrix, calc, positioning, arrows,shapes,backgrounds, arrows.meta, quotes}
\usepackage{gensymb,stmaryrd,mathtools,textcomp,xspace,grffile}
\usetikzlibrary{shapes.geometric,fit,matrix,positioning,shapes.multipart}
\usetikzlibrary{decorations.markings}
\usetikzlibrary{shadows,decorations.pathreplacing,fadings}
\pgfdeclarelayer{background}
\pgfsetlayers{background,main}
\usetikzlibrary{topaths, calc,3d}
\usetikzlibrary{fit}
\usepackage{color}

\usepackage{bm}		
\usepackage{pdflscape}	
\usepackage{ae,aecompl}
\usepackage[section]{placeins}
\usepackage{float}
\usepackage[normalem]{ulem}
\usepackage[toc,page]{appendix}
\usepackage{longtable}
\usepackage{adjustbox}

\usepackage[normalem]{ulem}
\usepackage{bbm}

\usepackage{titlesec}

\usepackage{scalerel}
\usepackage{tikz}
\usetikzlibrary{svg.path}

\definecolor{cacolor}{RGB}{51,0,102}

\definecolor{jccolor}{RGB}{66,151,0}

\definecolor{zlcolor}{RGB}{200,50,0}

\definecolor{bncolor}{RGB}{0,50,255}

\definecolor{kwcolor}{RGB}{106,90,205}

\definecolor{ndhcolor}{RGB}{204,102,0}

\definecolor{orcidlogocol}{HTML}{A6CE39}
\tikzset{
  orcidlogo/.pic={
    \fill[orcidlogocol] svg{M256,128c0,70.7-57.3,128-128,128C57.3,256,0,198.7,0,128C0,57.3,57.3,0,128,0C198.7,0,256,57.3,256,128z};
    \fill[white] svg{M86.3,186.2H70.9V79.1h15.4v48.4V186.2z}
                 svg{M108.9,79.1h41.6c39.6,0,57,28.3,57,53.6c0,27.5-21.5,53.6-56.8,53.6h-41.8V79.1z M124.3,172.4h24.5c34.9,0,42.9-26.5,42.9-39.7c0-21.5-13.7-39.7-43.7-39.7h-23.7V172.4z}
                 svg{M88.7,56.8c0,5.5-4.5,10.1-10.1,10.1c-5.6,0-10.1-4.6-10.1-10.1c0-5.6,4.5-10.1,10.1-10.1C84.2,46.7,88.7,51.3,88.7,56.8z};
  }
}

\newcommand\orcidicon[1]{\href{https://orcid.org/#1}{\mbox{\scalerel*{
\begin{tikzpicture}[yscale=-1,transform shape]
\pic{orcidlogo};
\end{tikzpicture}
}{|}}}}
\def\lcdm{$\Lambda$CDM}
\def\ukcmb{\ensuremath{\mu\rm{K}_{\rm{CMB}}}}
\def\cnnabbreviation{CNN}

\def\muka{$\mu$K-arcmin}

\def \mysec {\S}
\def \myfig {Fig.~}

\def \mytbl {Table~}

\def \uchicago {Department of Astronomy and Astrophysics, University of Chicago, Chicago, IL 60637, USA}
\def \uiuccs {Department of Computer Science, University of Illinois at Urbana–Champaign, Urbana, IL 61801, USA}
\def \berkeley {Department of Physics, University of California, Berkeley, CA 94720, USA}
\def \kicp {Kavli Institute for Cosmological Physics, University of Chicago, Chicago, IL 60637, USA}
\def \fnal {Fermi National Accelerator Laboratory, P.O. Box 500, Batavia, IL 60510, USA}
\def \umichigan {Department of Physics, University of Michigan, Ann Arbor, MI 48109, USA}
\def \umlctp{Leinweber Center for Theoretical Physics, University of Michigan, Ann Arbor, MI 48109, USA}
\def \mit {CSAIL, Massachusetts Institute of Technology (MIT), Cambridge, MA 02139, USA}

\def \slac {SLAC National Accelerator Laboratory \& KIPAC, 2575 Sand Hill Road, Menlo Park, CA 94025}

\def \figureImagesTrainingSampleExamplesPositiveNegative{
\begin{figure*}
\centering
\begin{subfigure}{0.98\textwidth}
	\includegraphics*[width=0.98\textwidth]{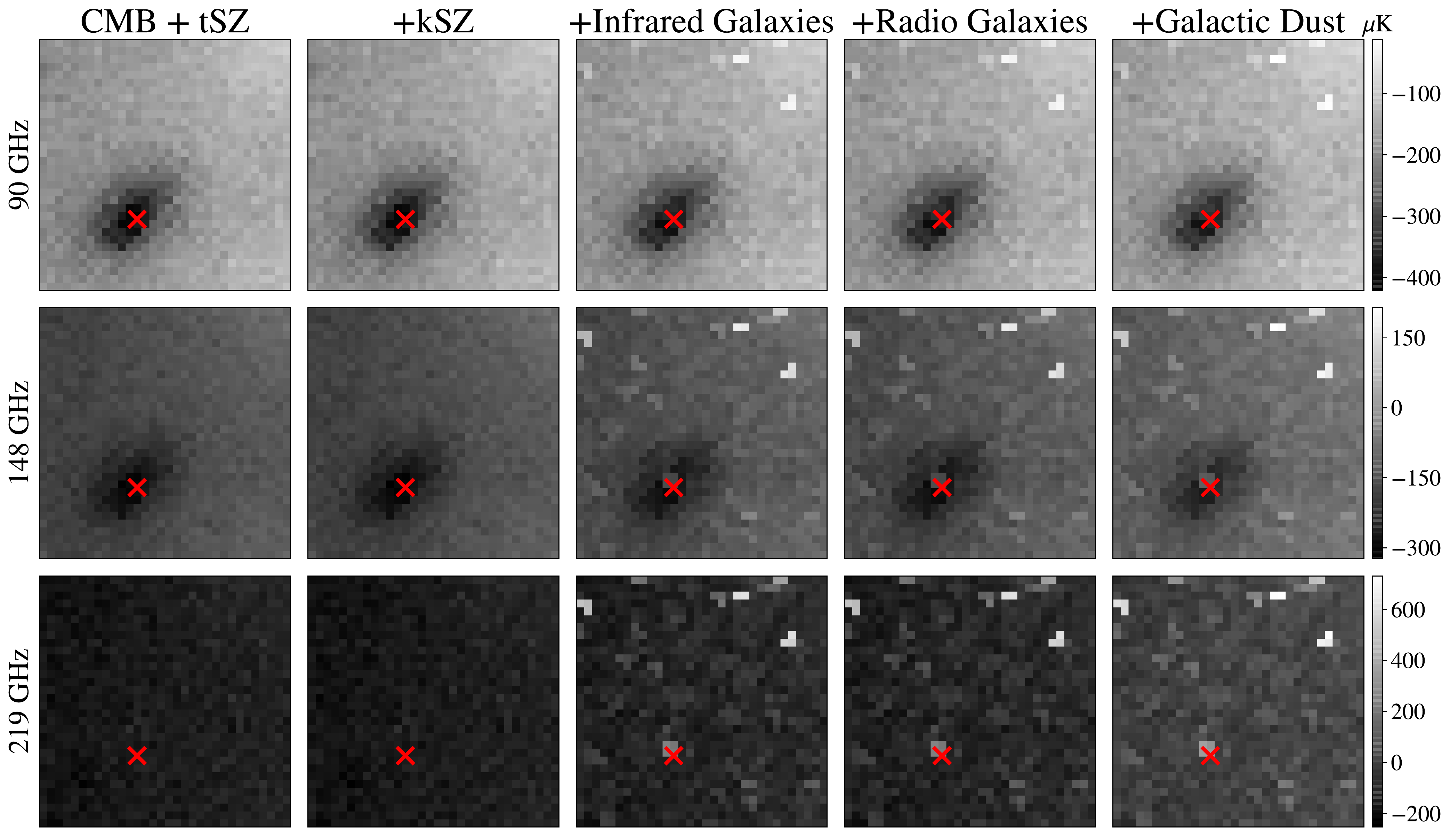}
   \caption{}
   \label{fig:inputPos1} 
\end{subfigure}
\begin{subfigure}{0.98\textwidth}
	\includegraphics*[width=0.98\textwidth]{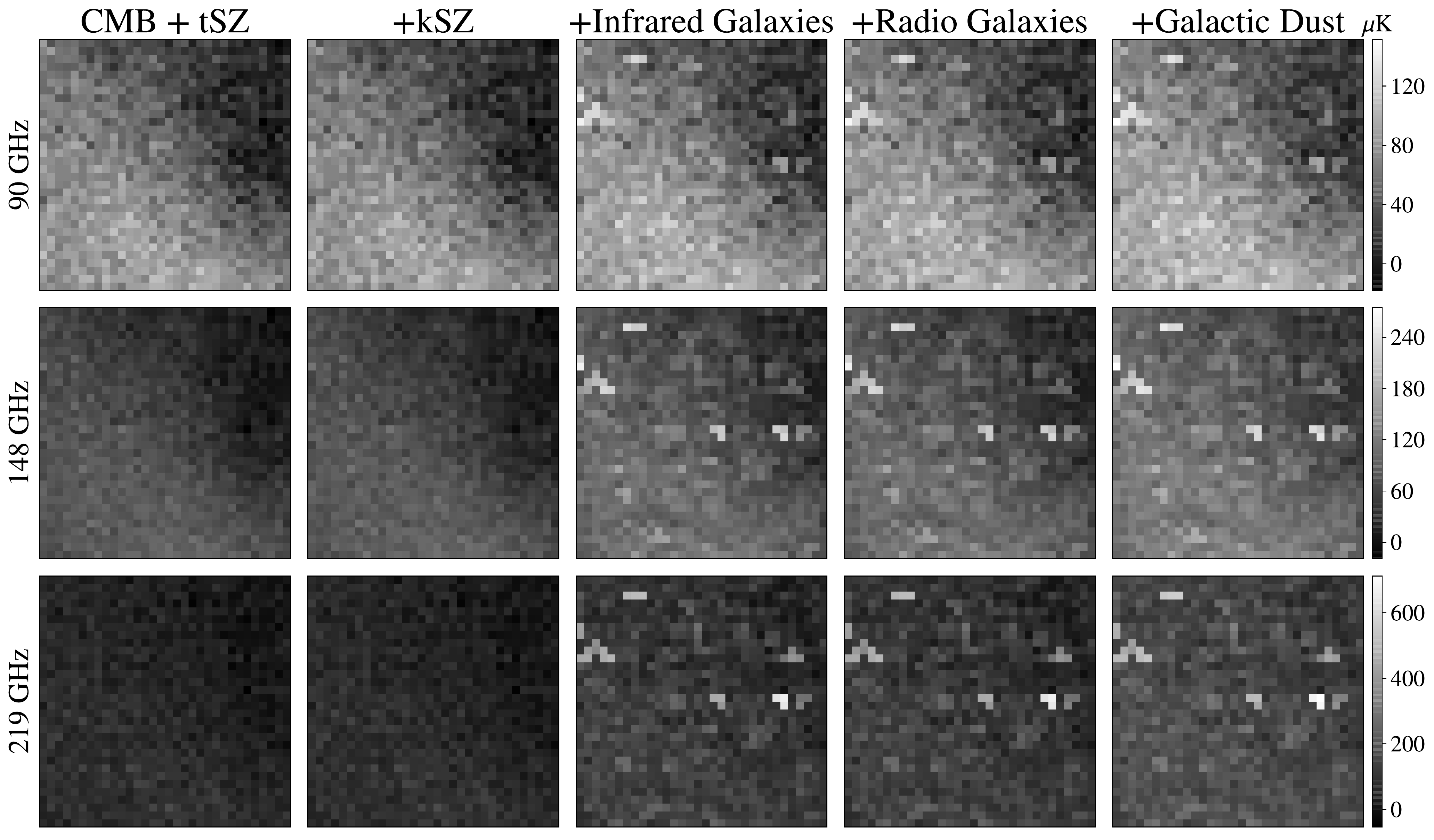}
   \caption{}
   \label{fig:inputNeg1}
\end{subfigure}
\caption[Loss and Accuracy]{
A positive (top) and negative (bottom) sample input to the neural network with three channels at 90GHz (1st row in top/bottom), 148 GHz (2nd row) and 219 GHz maps (3rd row). 
We show the cutout with different components (CMB, tSZ, kSZ, IR galaxies, radio galaxis, galactic dust) gradually added here. 
All cutouts contain instrument noise, which is 2.8 \muka\, 2.6 \muka\ and 6.6 \muka\ for 90 GHz, 148 GHz and 219 GHz maps respectively, consistent with projected performance for the upcoming CMB experiment SPT-3G.
Each pixel is 0.25 x 0.25 arcmin. Each cutout has 32x32 pixels.
  X-axis is ra, and Y-axis is dec. 
  The position of the cluster is market with a red "X". 
  }
\label{fig:ImagesTrainingSampleExamplesPositiveNegative}
\end{figure*}
}

\def \figureImagesTrainingSampleExamplesPositiveNegativeInfrared{
\begin{figure}
  \centering
	\includegraphics*[width=0.4\textwidth]{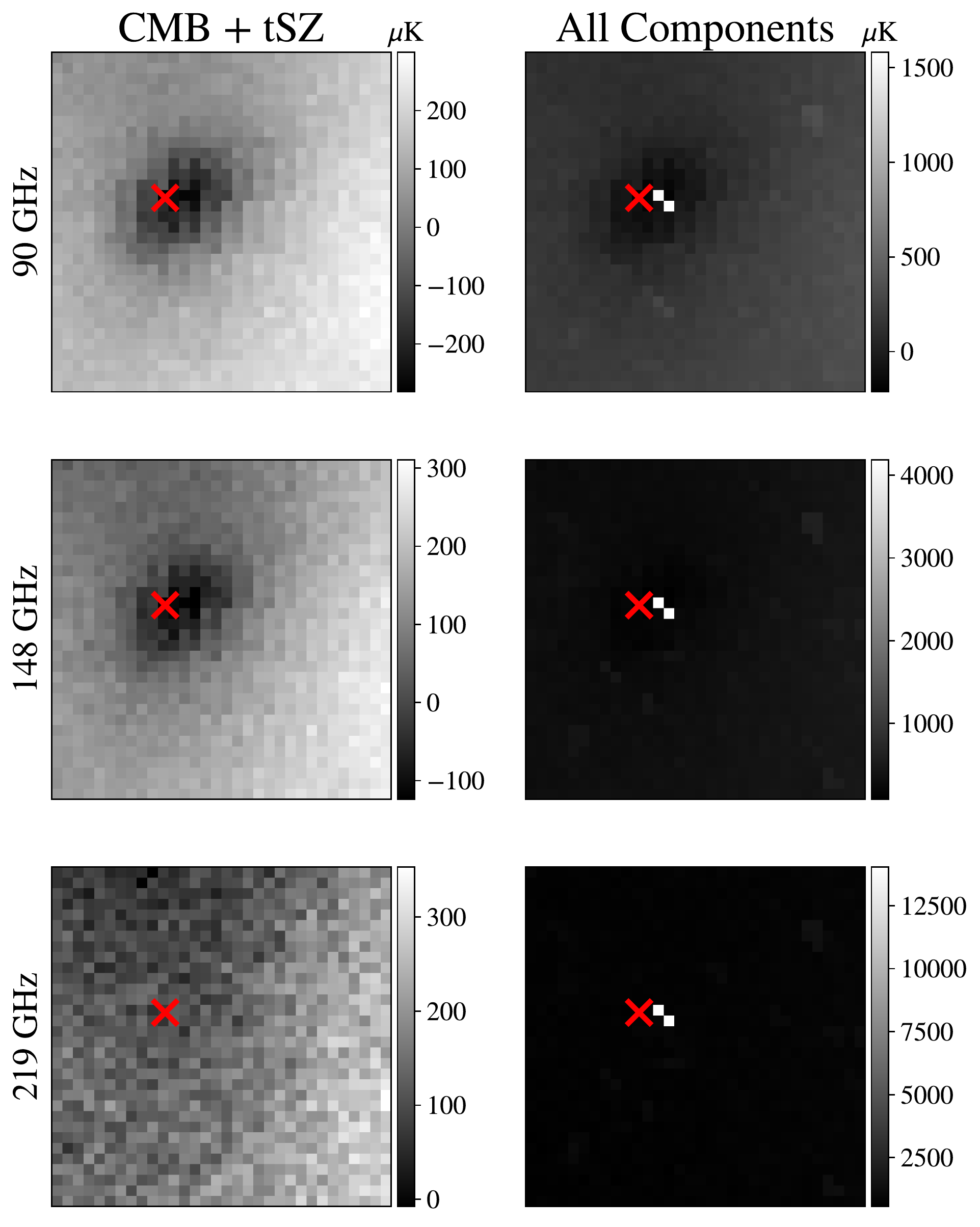}
  \caption{We provide an example cutout that contains a cluster, and also happens to host a high-flux infrared galaxy.  The column on the left is the image with only the CMB$+$tSZ components, and the column on the right is the image with all components adding on the kSZ, infrared and radio galaxies, and galactic dust.  Similar to \myfig\ref{fig:ImagesTrainingSampleExamplesPositiveNegative}, top to bottom rows correspond to the 90, 148, and 219 GHz bands.  The addition of a high-flux infrared galaxy increases the colorbar range by an order of magnitude, weakening the tSZ signal from the galaxy cluster relative to the rest of the cutout.  The CNN takes input images, such as those in the right-hand column, with no pre-processing.
  \label{fig:ImagesTrainingSampleExamplesPositiveNegativeInfrared}}
\end{figure}
}

\def \figureNeuralNetworkTrainingHistory{
\begin{figure*}
\centering
\begin{subfigure}{0.8\textwidth}
	\includegraphics*[width=0.95\textwidth]{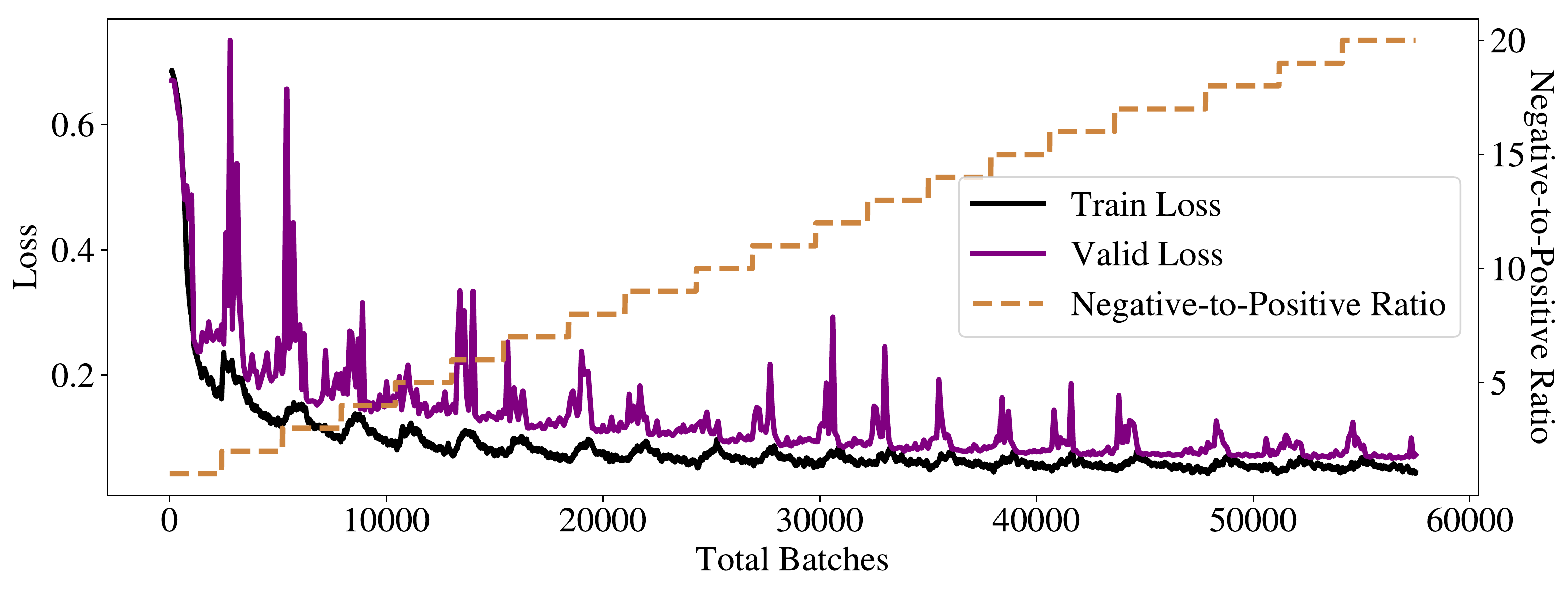}
   \caption{}
   \label{fig:Ng1} 
\end{subfigure}
\begin{subfigure}{0.8\textwidth}
	\includegraphics*[width=0.95\textwidth]{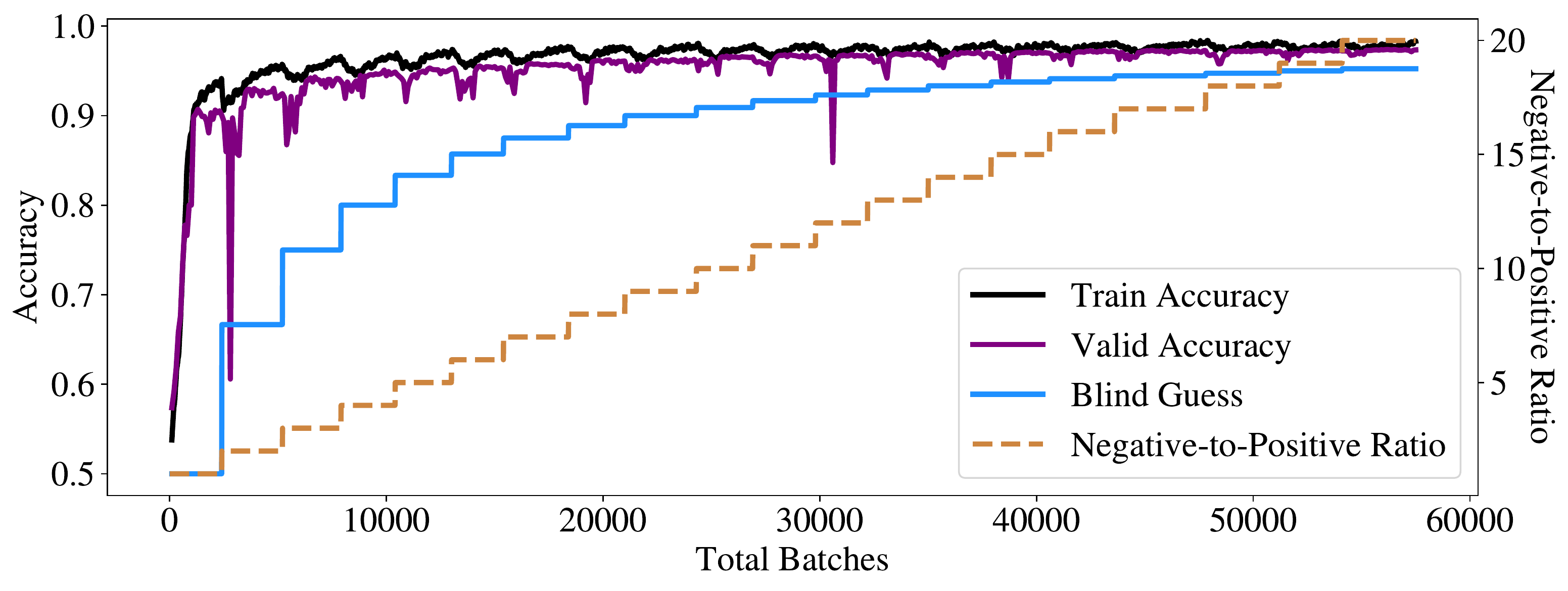}
   \caption{}
   \label{fig:Ng2}
\end{subfigure}
\caption[Loss and Accuracy]{In this figure, (a) shows the loss history, and (b) shows the accuracy history as a function of the number of batches (iterations).}
\label{fig:NeuralNetworkTrainingHistory}
\end{figure*}
}

\def \figureImagesPredictionExamples{
\begin{figure}
  \centering
	\includegraphics*[width=0.45\textwidth]{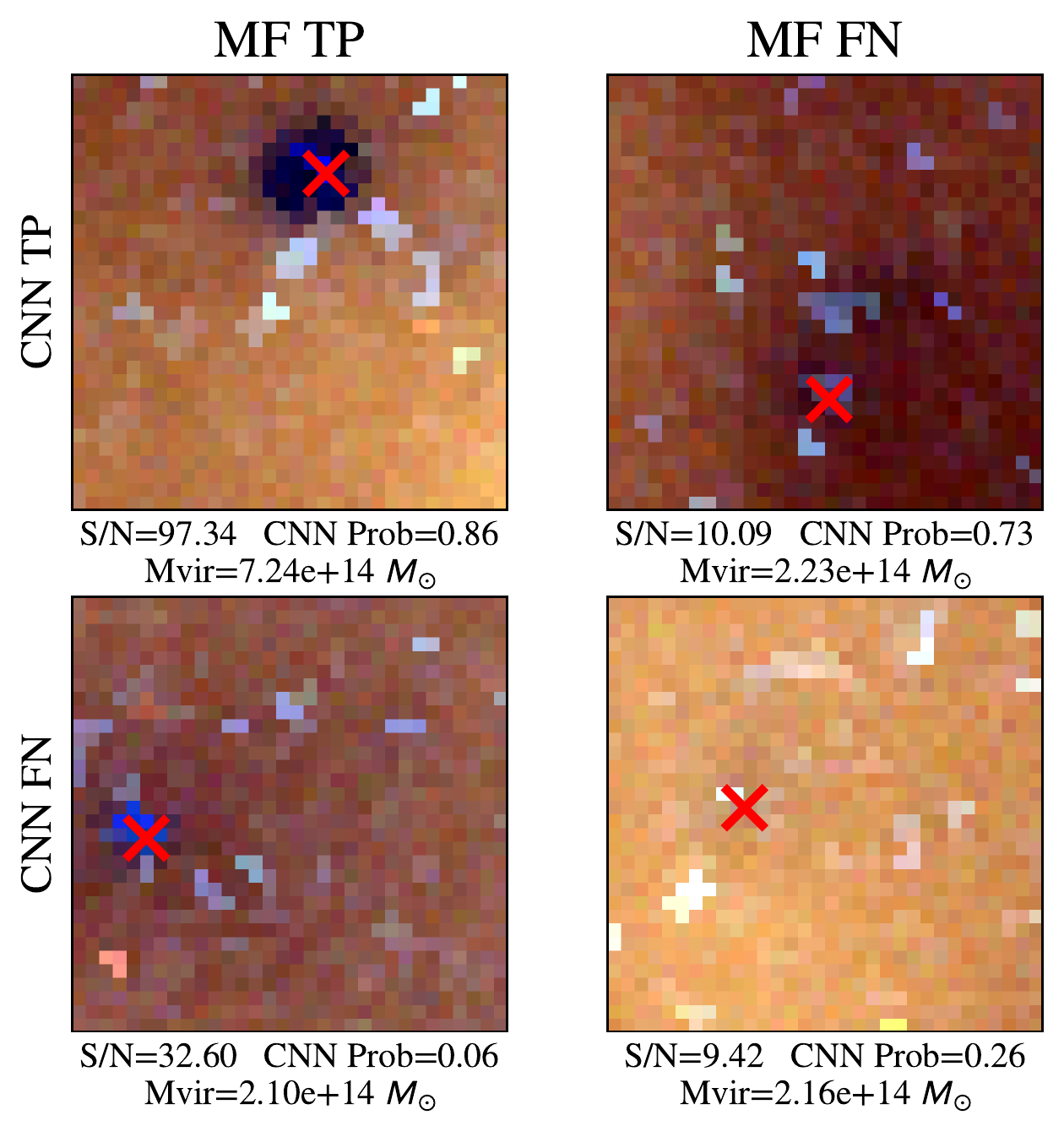}
  \caption{
  Examples of cluster-containing cutouts correctly classified by the combined method EnsemblePROD. 
  The pictures are generated by treating the three channels (90/148/219 GHz) of input as RGB channels. For each channel, the pixel value is re-scaled to [0, 255], with the same channel (e.g. red channel) of all cutouts sharing the same color scale.
  The top (bottom) row has high (low) probabilities assigned by the CNN classifier, which we identify as a CNN "true positive"/TP (CNN "false negative"/FN).  The left (right) column has high (low) signal to noise as measured by the MF algorithm, which we identify as a MF "true positive", MF TP (MF "false negative", MF FN).  
  While the bottom right panel might be identified as both a CNN FN and a MF FN, the EnsemblePROD correctly identifies this cutout because of its relatively high ranking in the CNN and MF scores. 
  Note that although the score given by CNN in the bottom left is only 0.06, we can see that it is a  high percentile already because CNN gives most cutouts a score of essentially 0 (see the histogram in \myfig\ref{fig:FoneScoreComparison})
  \label{fig:EPROD_MF_CNN_TPFN}}
\end{figure}

}

\def \figureCompletenessMvir{
\begin{figure*}
  \centering
	\includegraphics*[width=0.9\textwidth]{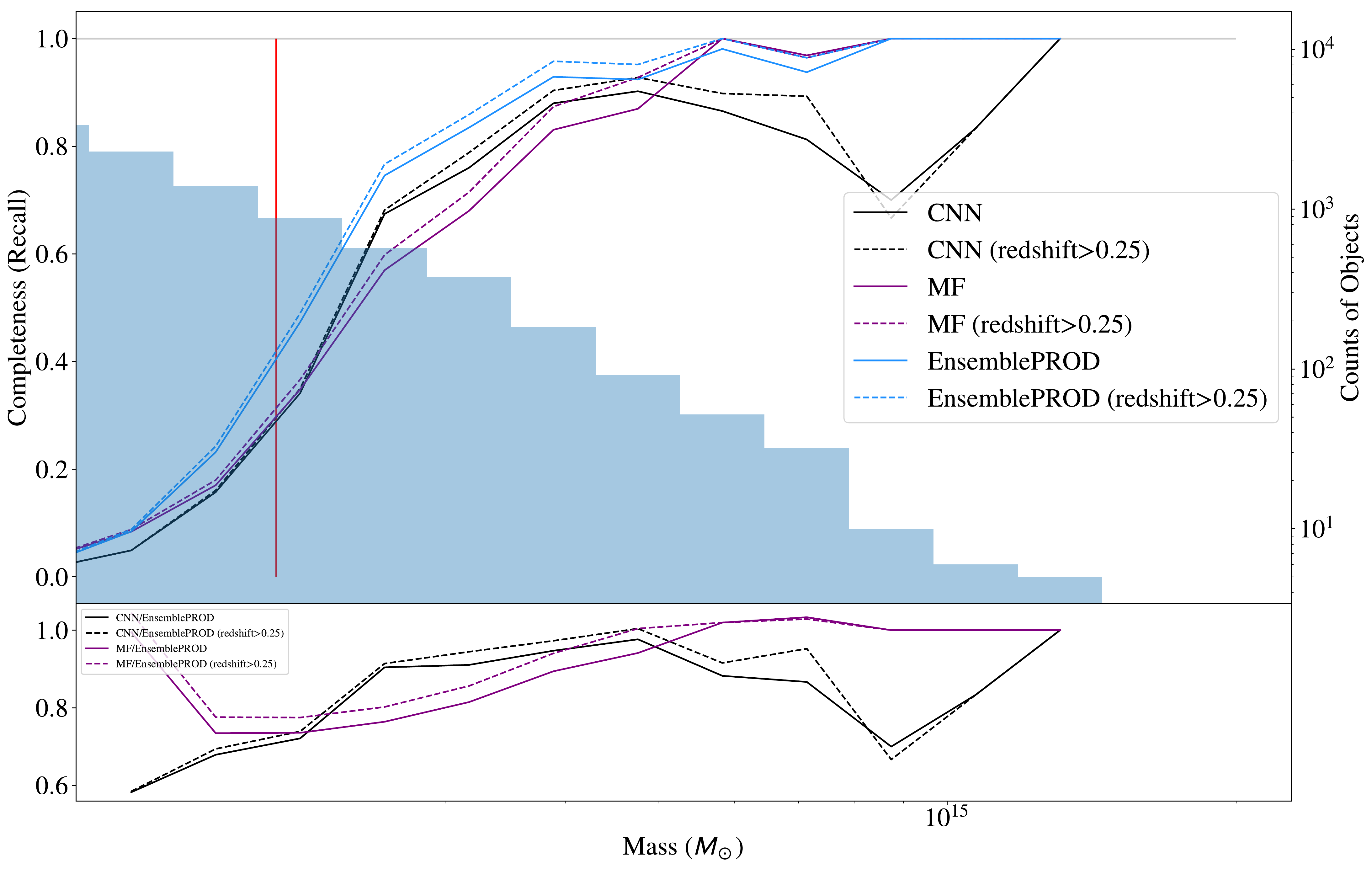}
  \caption{Top panel: Blue histogram shows underlying mass function of the simulated galaxy cluster sample in the cutouts (Counts of Objects vs. Mass).  
  Lines show the completeness curves for each method, with the dashed lines corresponding to the completeness curve for objects that satisfy our redshift threshold of $z>0.25$ used for training. The purple dashed lines correspond to the MF, black dashed the CNN, and blue dashed the EnsemblePROD. We can see that the EnsemblePROD completeness curve sits above both methods until $\sim4\times10^{14}$M$_\odot$, where there are few objects.  The decrease of the CNN curve at the highest masses could be due to the decreased sample size for training in the corresponding mass range. 
  Bottom panel: Ratio of CNN (or MF) completeness curve to the corresponding EnsemblePROD completeness curve with the same line style and color as the top panel.    
  We also include solid lines for the full sample without the redshift threshold, but since the impact of the redshift threshold is small, they are very similar.
  \label{fig:CompletenessMvir}}
\end{figure*}
}

\def \figureCompletenessRedshift{
\begin{figure*}
  \centering
	\includegraphics*[width=0.9\textwidth]{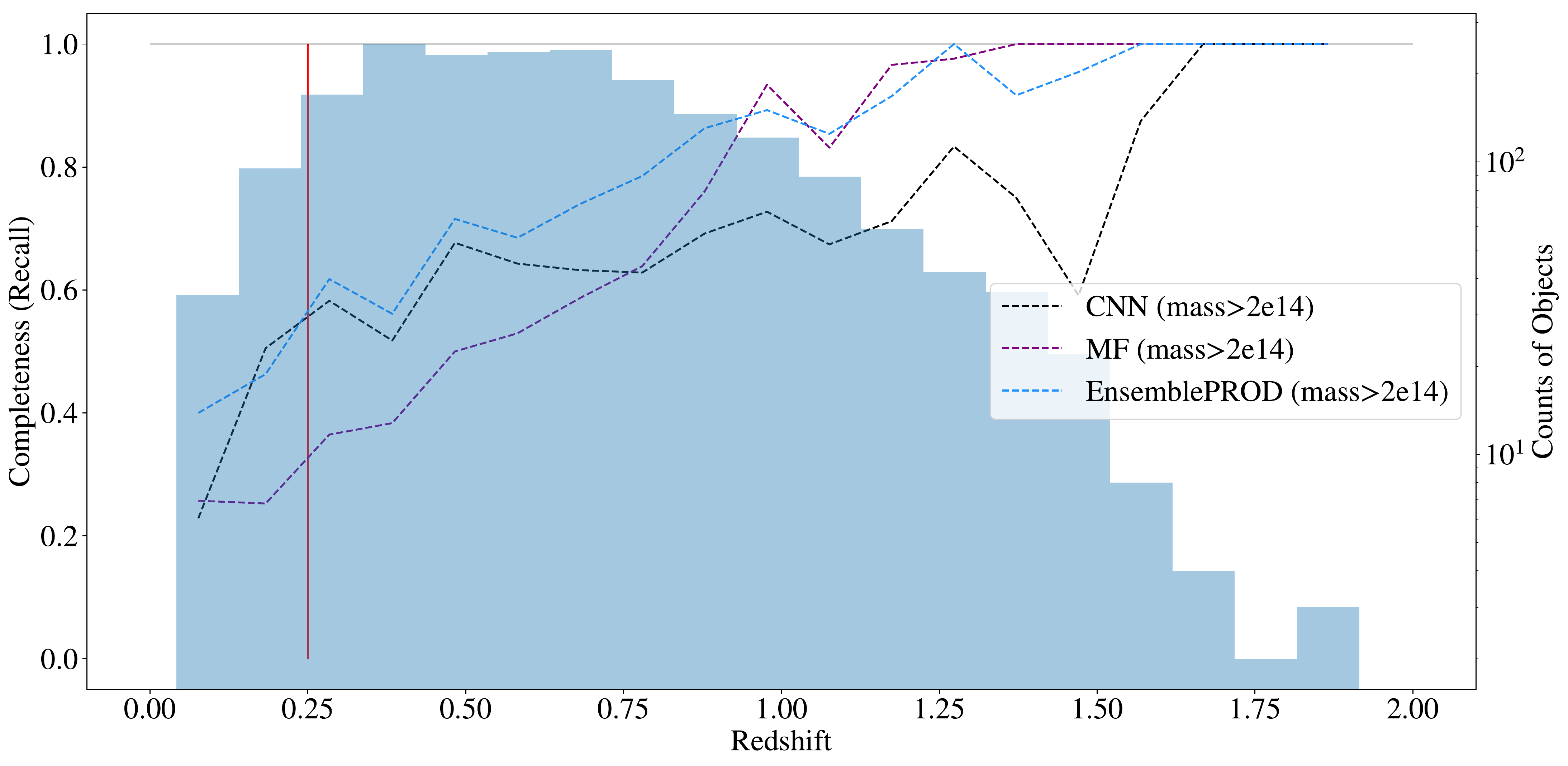}
  \caption{Blue histogram shows the underlying number of galaxy clusters in each redshift bin that satisfy our mass threshold of $M_{halo}>2\times10^{14}$M$_\odot$ used for training (Counts of Objects vs. Redshift).  Lines show the completeness curves for each method on the subset of objects over the same mass threshold.  The EnsemblePROD completeness curve sits above the other methods until $z\approx0.9$, at which point it is comparable to MF. 
  \label{fig:CompletenessRedshift}}
\end{figure*}
}

\def \figureFoneScoreComparison{
\begin{figure}
  \centering
	\includegraphics*[width=0.48\textwidth]{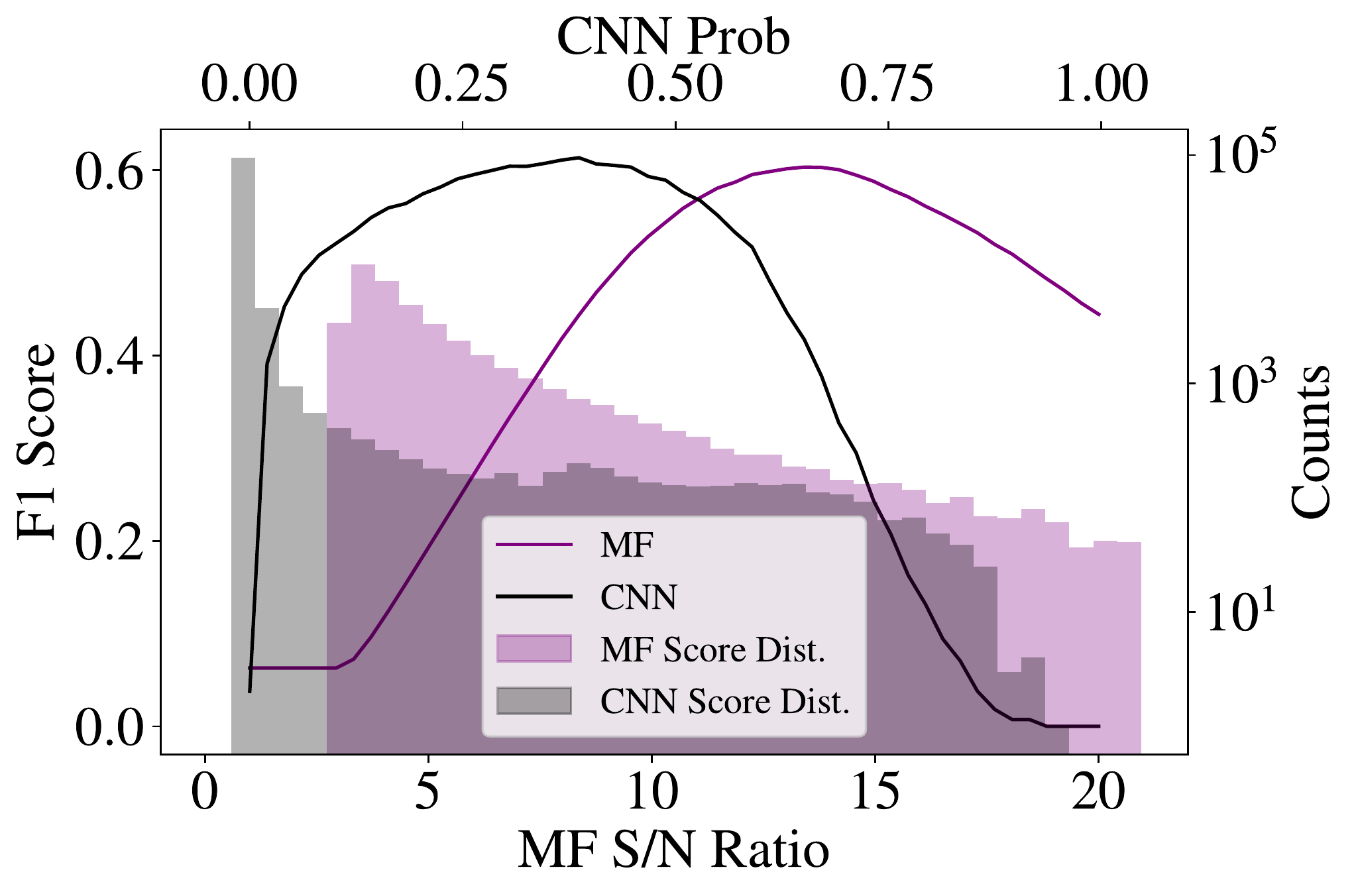}
  \caption{F1 score as a function of the threshold for S/N ratio from MF and probability from CNN (on the validation set). 
  We pick the thresholds for each method here and later evaluate them on the test sets.
  The distribution of the scores (S/N ratio for MF, and probability for CNN) are shown in the background. 
  We also did not include any MF SNRs below 3.0 because there are too many of them.
  \label{fig:FoneScoreComparison} 
  }
  
\end{figure}
}

\def \figureFOneScoreEnsembleMethods{
\begin{figure*}
\centering
\begin{subfigure}[h]{0.45\linewidth}
	\includegraphics[width=0.95\textwidth]{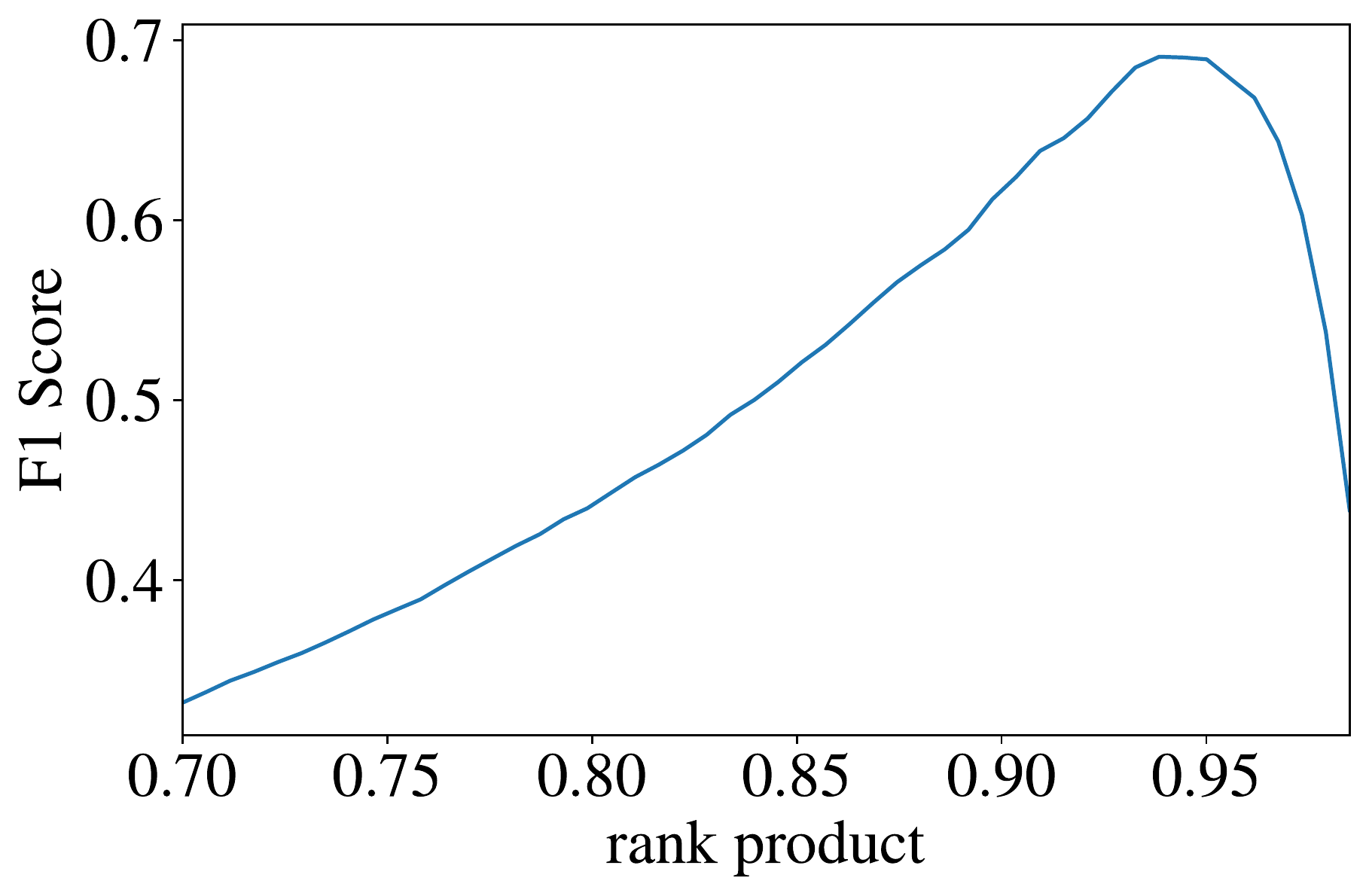}
  \caption{F1 score on the validation set if we form a single score by multiplying the rank percentile of CNN score and MF S/N ratio. 
  The validation set selects the rank product threshold to be 93.8\%.
  }
  \label{fig:FOneScoreEnsemblePROD} 
\end{subfigure}
\hfill
\begin{subfigure}[h]{0.45\linewidth}
	\includegraphics[width=0.95\textwidth]{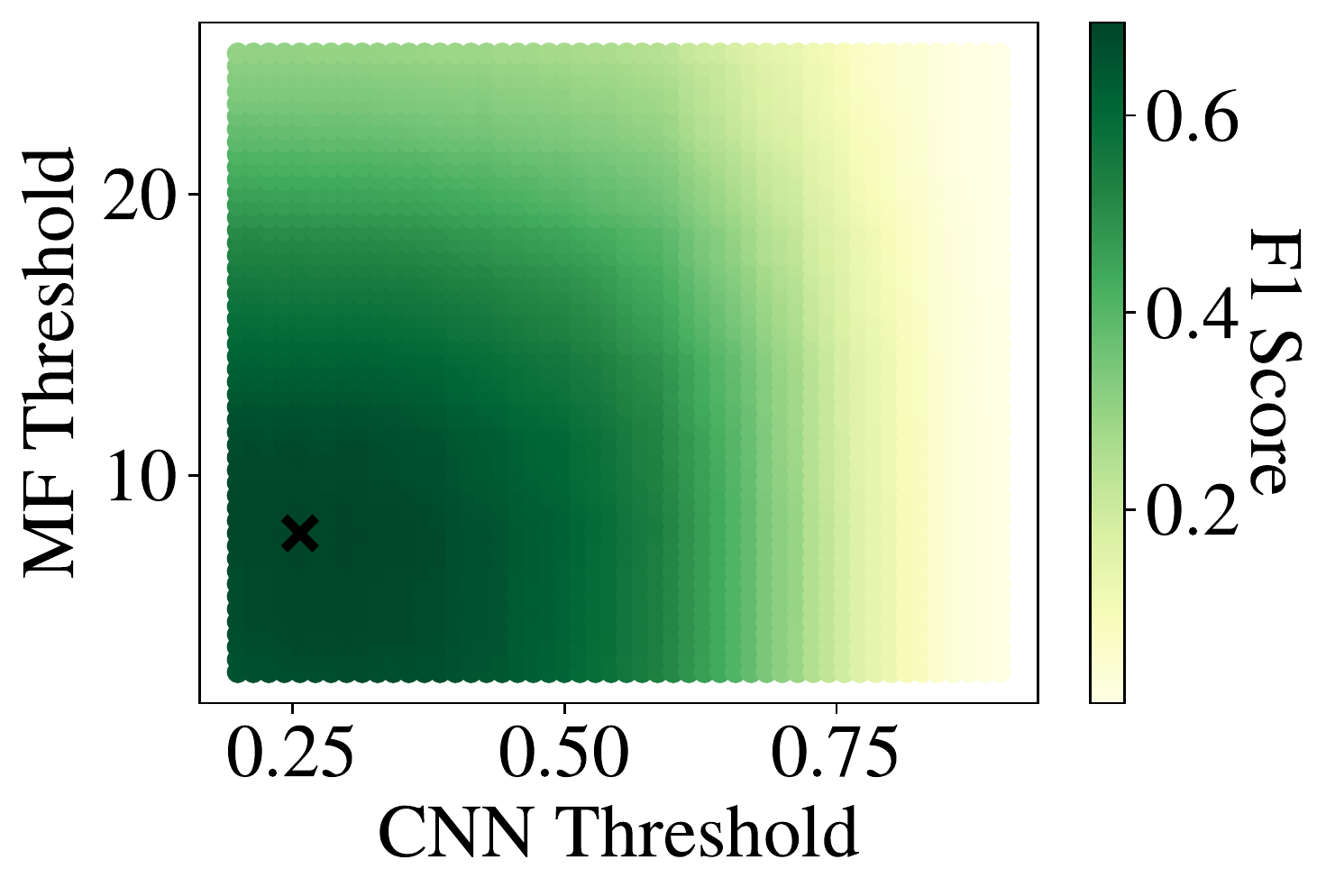}
  \caption{F1 score on the validation set if we require both MF and CNN's score to pass a certain threshold. 
  The optimal point is when CNN probability is greater than 0.26 and MF S/N ratio greater than 7.9 (marked with a black cross).}
  \label{fig:FOneScoreEnsembleAND}
\end{subfigure}

\caption[(Left) The F1 score for the Ensemble product..... (Right)  The F1 score for the EnsembleAND..... F1 Scores for Ensemble Methods]{Here we show the performance of 2 Ensemble methods: EnsemblePROD (\ref{fig:FOneScoreEnsemblePROD}) and EnsembleAND (\ref{fig:FOneScoreEnsembleAND}). 
Their performances are very similar: Both achieve an F1 score of 0.69 on the validation set, and \textbf{0.68} on the test set.
Both methods, however, noticeably outperform CNN or MF standalone. 
}
\label{fig:FOneScoreEnsembleMethods}
\end{figure*}
}

\def \figureNeuralNetworkPerformanceMetrics{
\begin{figure}
\centering
\begin{subfigure}[h]{\linewidth}
	\includegraphics*[width=\linewidth]{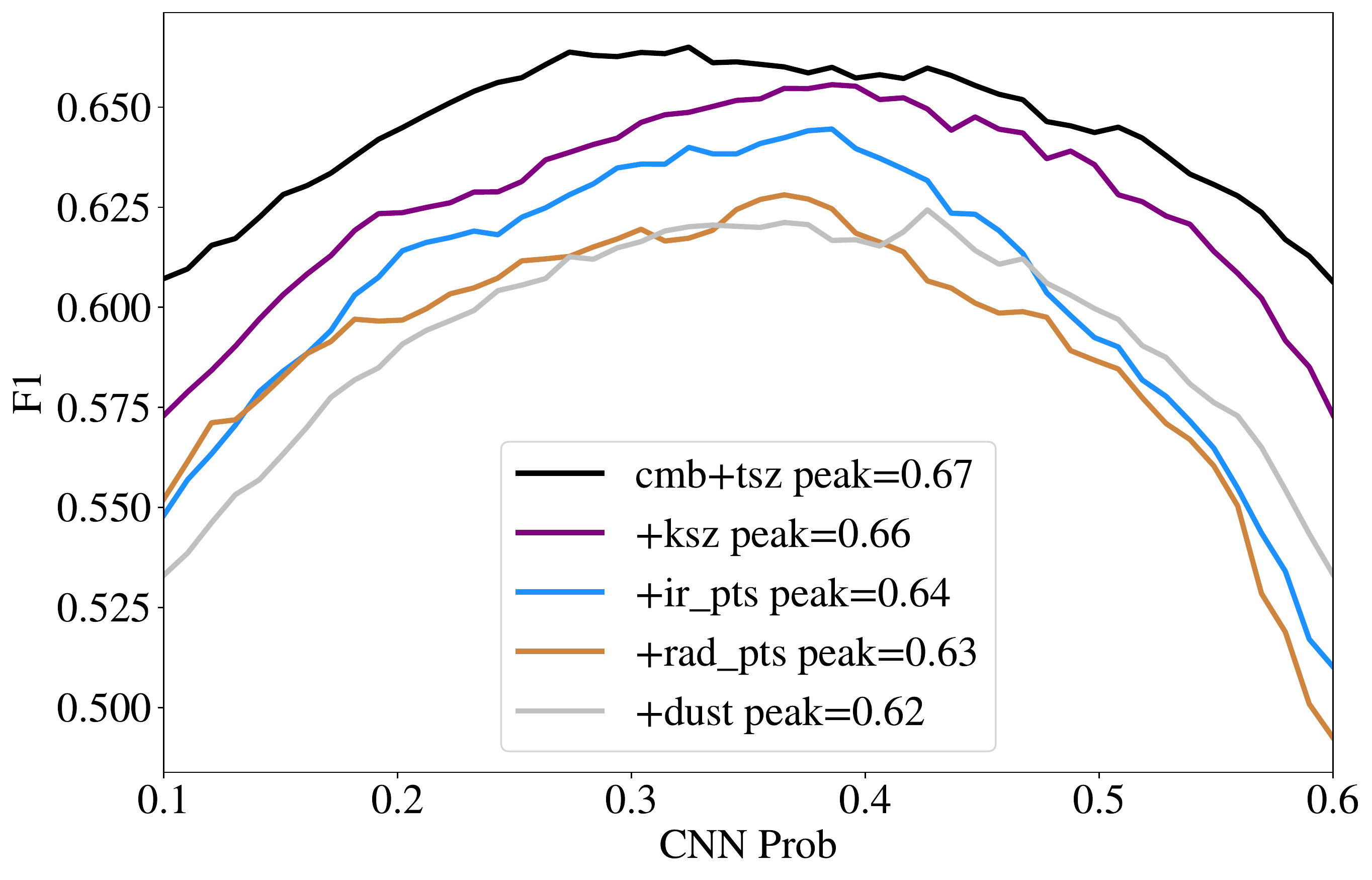}
  \caption{}
  \label{fig:NeuralNetworkPerformanceMetricsF1} 
\end{subfigure}
\vfill
\begin{subfigure}[h]{\linewidth}
	\includegraphics*[width=\linewidth]{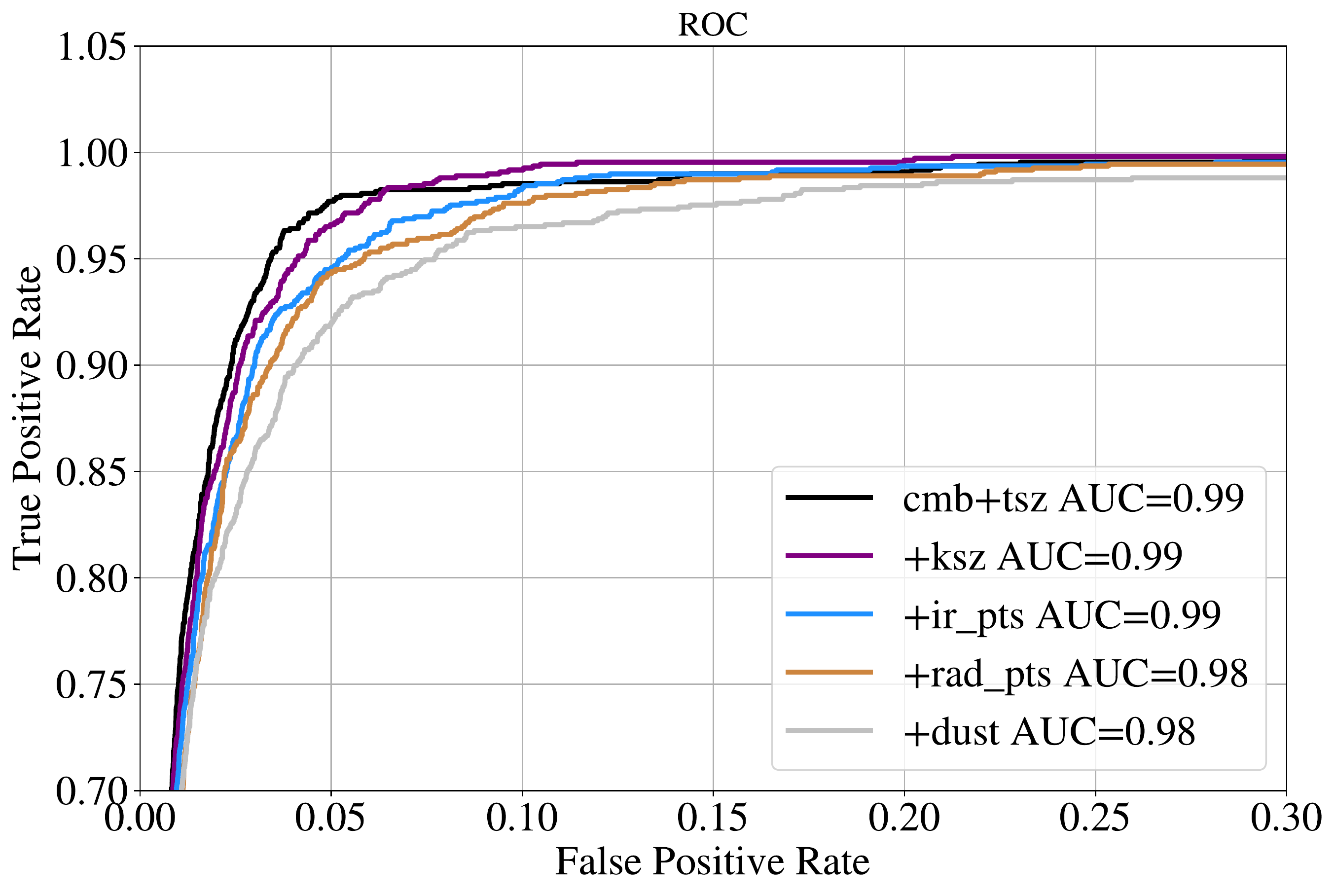}
  \caption{}
  \label{fig:NeuralNetworkPerformanceMetricsROC}
\end{subfigure}

\caption[F1 and ROC of CNN with differnet components added]{Metrics: \ref{fig:NeuralNetworkPerformanceMetricsF1} shows the curves of F1 score as a function of the threshold on \textbf{CNN probability} for different levels of noises. \ref{fig:NeuralNetworkPerformanceMetricsROC} shows the ROCs of CNN's predictions on different levels of noises. 
}
\label{fig:NeuralNetworkPerformanceMetrics}
\end{figure}
}

\def \figureSelFuncBoth{
\begin{figure*}
  \centering
	\includegraphics*[width=0.9\textwidth]{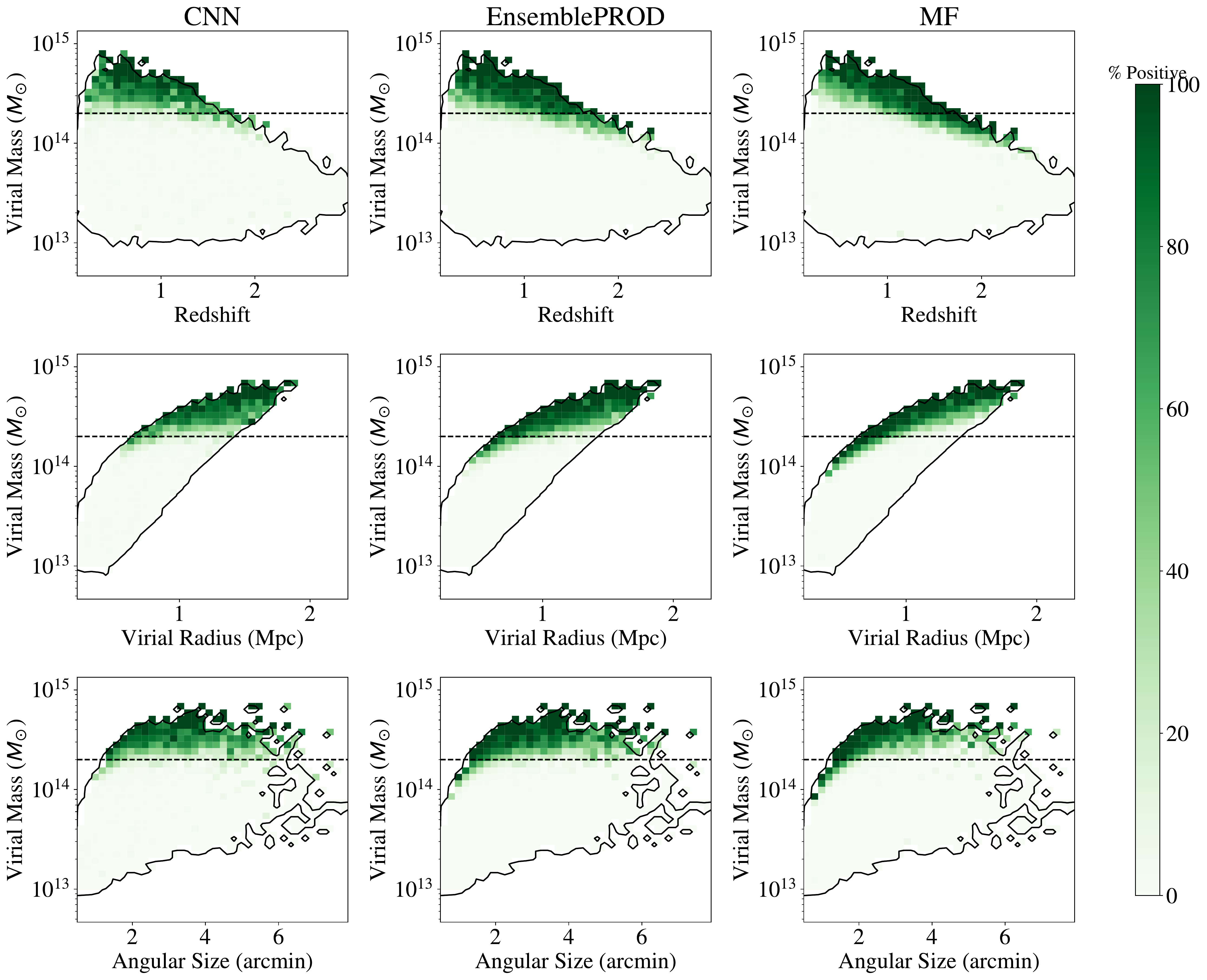}
  \caption{We plot the distribution of cluster properties and color code by true positive identification of each method.  The left column corresponds to true positive identification by our CNN, center by the rank product, and right by MF.  
  The black dashed line illustrates our chosen mass cut for what we labeled as true positives for the training set in our CNN.  The color coding of \% Positive colors the percent of clusters (cutout containing clusters) in that cluster property bin that was positively identified by each method.  The property bins have at least 3 clusters from which we calculate a percentage.  One key feature to note is that the CNN method has a relatively stronger mass-limited selection function that is driven by our mass cut in labeling true positives.  
  This selection does not strongly depend on redshift, virial radius, or angular size.  On the other hand, the matched filter preferentially picks out lower mass clusters that are more compact, which are those at higher redshifts.  The combined method, using the rank product, also has a selection function more aligned with our chosen mass cut.  
  \label{fig:split2_selfunc_both}}
\end{figure*}
}

\def \figureFailuresAndShifted{
\begin{figure*}
\centering

\begin{subfigure}[b]{0.9\textwidth}
    \centering
\includegraphics*[width=1.0\textwidth]{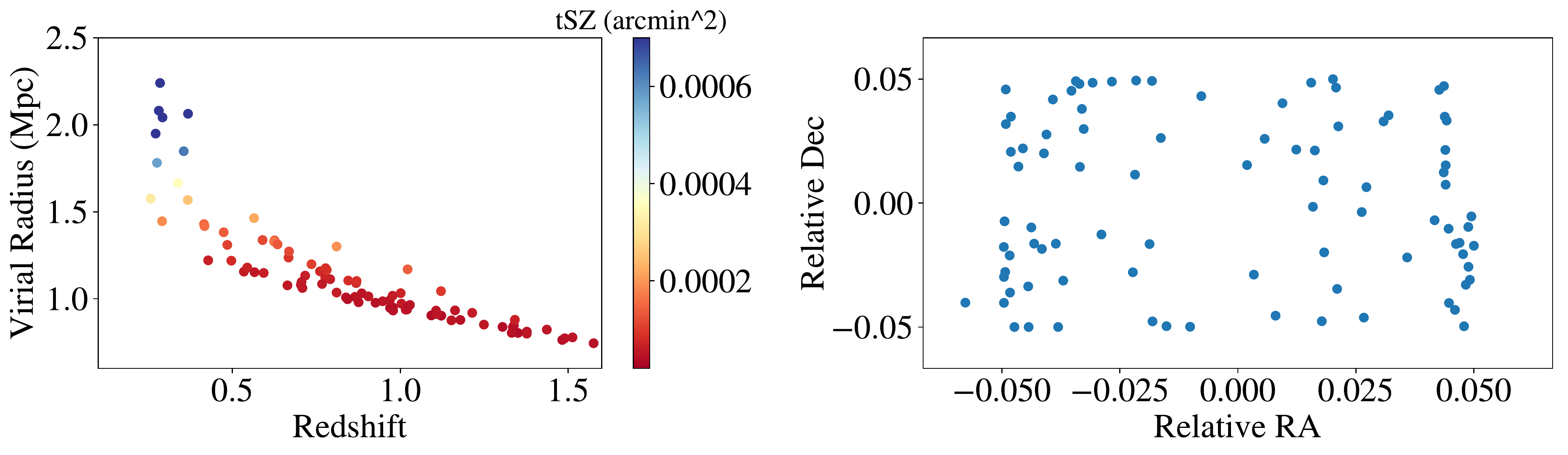}
\caption{ (Left) The virial radius as a function of redshift colorcoded by their tSZ of 88 clusters with a high MF score, but low ranked score from the CNN. (Right) Position of the clusters in the cutout shown on the right.  These clusters span the mass and redshift range, but tend to sit close to the edges of the cutouts that the CNN evaluated.
\label{fig:split2_CNNfailures}
   }
\end{subfigure}

\begin{subfigure}[b]{0.9\textwidth}
 	\includegraphics*[width=\textwidth]{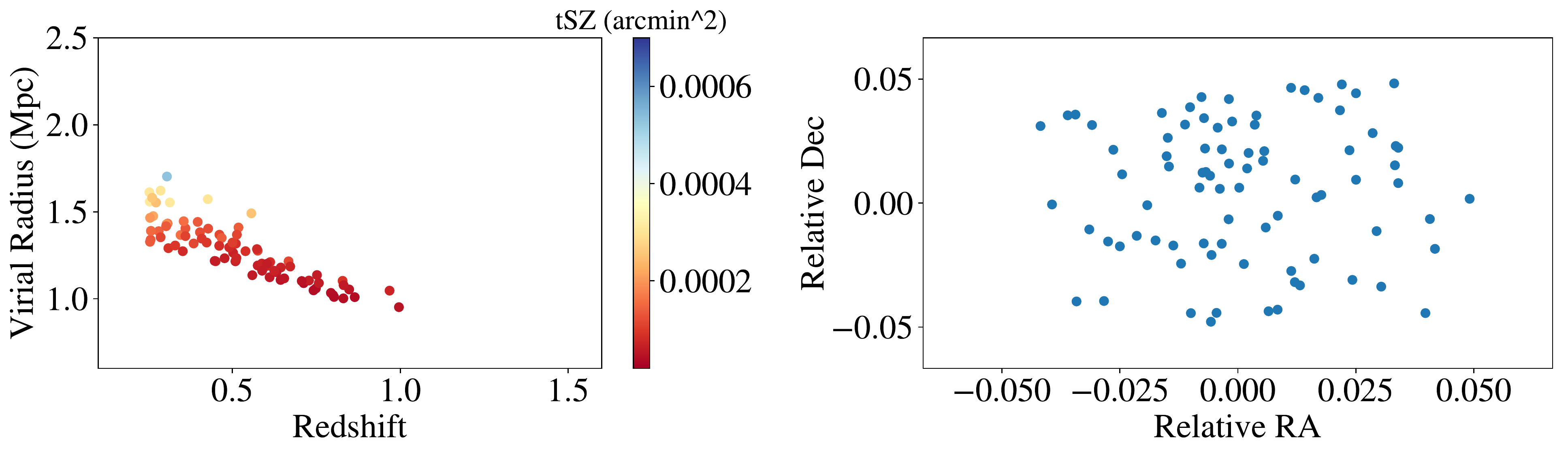}
   \caption{(Left) The virial radius as a function of redshift color-coded by their tSZ of 73 clusters with a high CNN score, but low MF S/N ratio. (Right) Position of the clusters in the cutout shown on the right.  
   \label{fig:split2_MFfailures}}
 \end{subfigure}
 
\begin{subfigure}[b]{1.0\textwidth}
\includegraphics[width=0.66\textwidth]{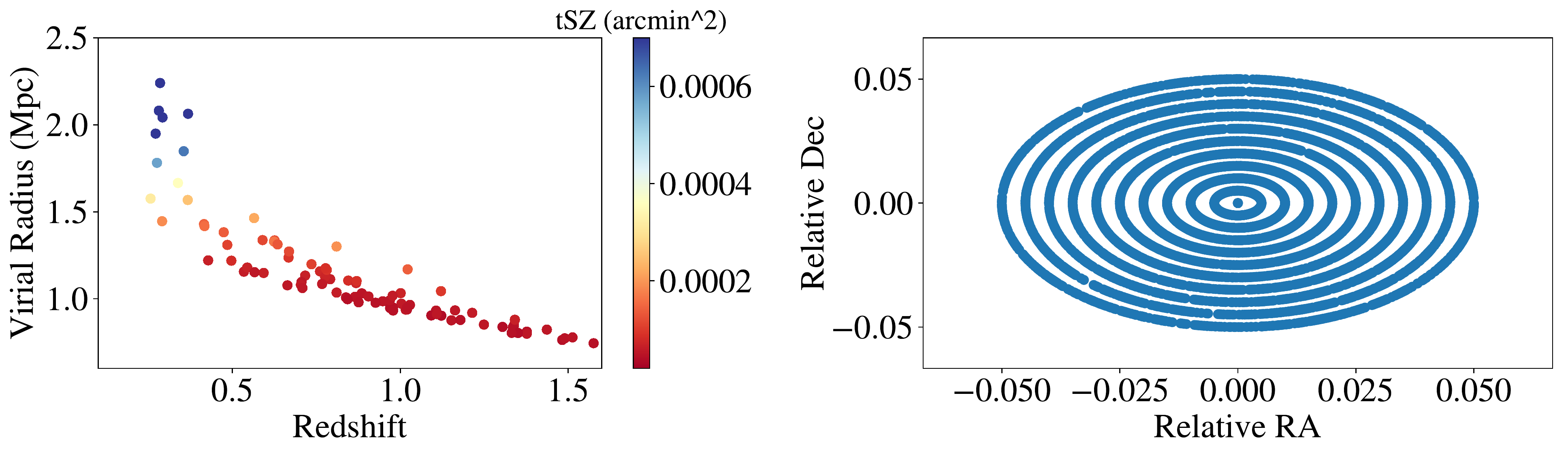}
\includegraphics[width=0.33\textwidth]{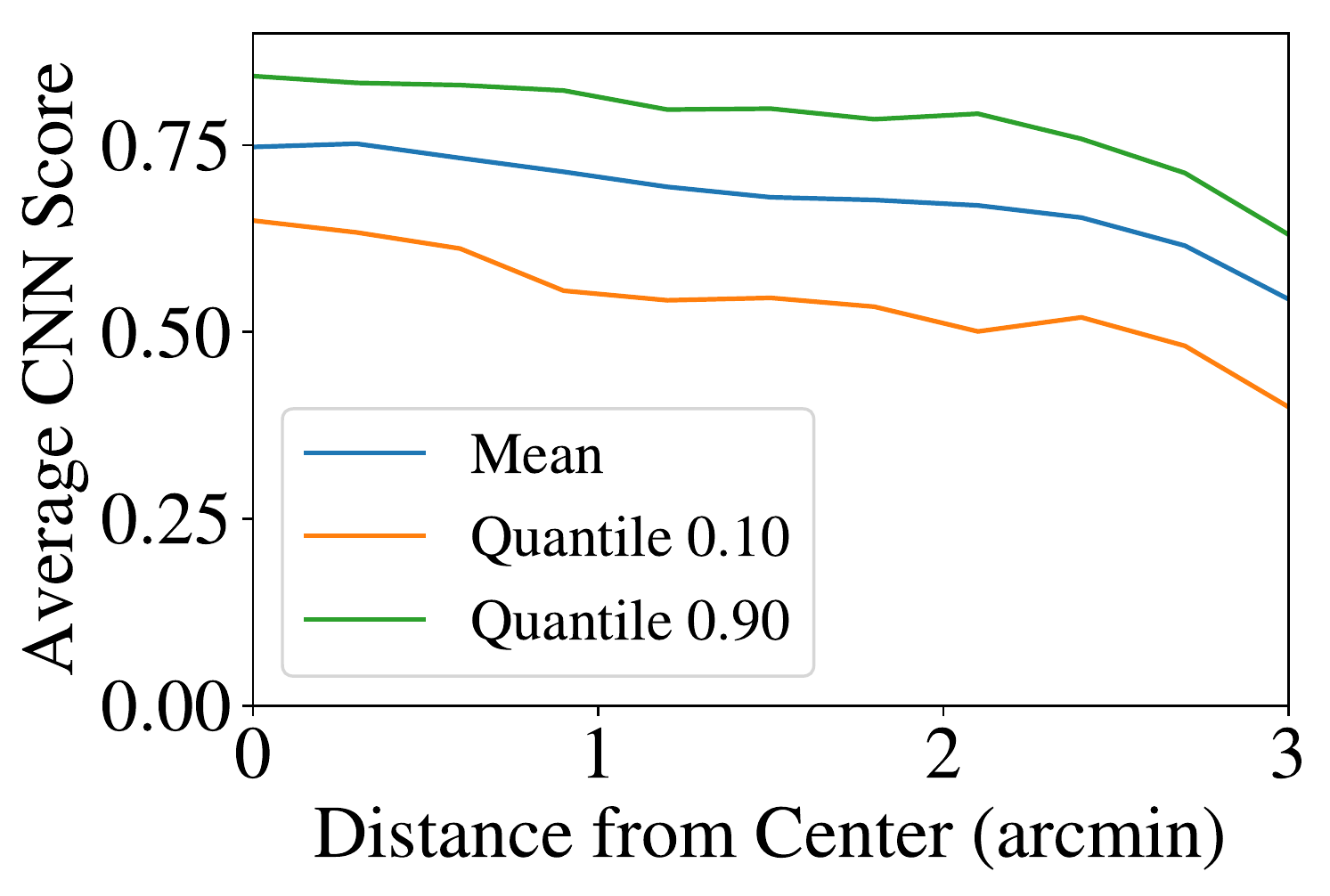}
\caption{
To investigate whether an edge effect is indeed the main driver of the low prediction probability on these clusters, we randomly regenerated cutouts with these clusters located at varying distances from the center of the cutouts.
 On average, the prediction value given by CNN is a decreasing function of the distance between the center of the cluster and the center of the cutout. 
 We first divide the 88 clusters in \ref{fig:split2_CNNfailures} into 11 groups. 
 For each group, we randomly regenerate cutouts for these clusters such that the cluster has different distance from and angle relative to the cutout center (Middle), and then compute the curve of prediction score as a function of distance from center for these randomly generated cutouts.
 Finally, we plot the mean, 10th percentile and 90th percentile (over the 11 groups) of these curves
 \label{fig:split2_pred_over_dist}.
}
\end{subfigure}

\caption{}
\end{figure*}
}

\def \figureNeuralNetworkArchitecture{
\begin{figure*}
\centering
\begin{tikzpicture}[node distance=0.2cm,
simple/.style={draw},
emptyAnchor/.style={circle, fill=white}, 
ResStage/.style={outer sep=1pt, draw,text width=5cm,minimum height=2cm,minimum width=5cm},
ConvBN/.style={outer sep=1pt, draw,text width=3.5cm,minimum height=0.5cm,minimum width=3.5cm},
MaxPool/.style={ellipse, outer sep=1pt, draw, minimum height=0.1cm, minimum width=2cm},
plusReLU/.style={ellipse, fill=white, draw},
edgeReLU/.style={midway, right=4pt, fill=white},
myline/.style={thick,black!50},
shorter/.style={shorten <=1mm,shorten >=0.5mm},
]
\footnotesize
    \node[simple] (data_input) {data};
    \node[ConvBN] (conv11) [below=0.2cm of data_input]           {3x3 Conv-BN-ReLU\\ channels: 64};
    \node[ConvBN] (conv12) [below=of conv11]               {3x3 Conv(s=2)-BN-ReLU\\ channels: 64};
    \node[MaxPool] (pool1) [below=of conv12]               {Max Pool(s=2)};
    \node[ResStage, fill=yellow!30] (resstage1) [below=of pool1]           {Stage 1 (3 Blocks)\\ channels:\\ \hspace{2pt} (in,bottleneck,out) = (64,64,256)\\  \textcolor{red}{1st block stride=1}};
    \node[ResStage, fill=blue!30] (resstage2) [below=of resstage1]           {Stage 2 (4 Blocks)\\  channels:\\ \hspace{2pt} (in,bottleneck,out) = (256,128,512)\\ 1st block stride=2};
    \node[ResStage, fill=red!30] (resstage3) [below=of resstage2]           {Stage 3 (6 Blocks)\\ channels:\\ \hspace{2pt}   (in,bottleneck,out) = (512,256,1024)\\ 1st block stride=2};
    \node[ConvBN] (fc1) [below=of resstage3]               {FC-ReLU\\ channels: 256};
    \node[ConvBN] (fc2) [below=of fc1]               {FC-ReLU\\ channels: 256};
    \node[ConvBN] (fc3) [below=of fc2]               {FC-ReLU-Softmax\\ channels: 2};
    \node[ConvBN] (output) [below=of fc3]  {prediction:\\ ($\mathbb{P}$(positive), $\mathbb{P}$(negative))};
    \node[emptyAnchor] (resstage1top2) [above right = 1.2cm and 4.5cm of resstage1]               {}; 
    \node[simple] (resstage1input) [below= of resstage1top2]               {Stage 1 input};
    \node[ConvBN] (res2a_branch2a) [below right =0.5cm and -1cm of resstage1input]           {3x3 Conv\textcolor{red}{(s=1)}-BN-ReLU \\ channels: 64 (bottleneck)};
    \node[ConvBN] (res2a_branch2b) [below = of res2a_branch2a]           {3x3 Conv-BN-ReLU \\ channels: 64 (bottleneck)};
    \node[ConvBN] (res2a_branch2c) [below = of res2a_branch2b]           {3x3 Conv-BN\\ channels: 256};
    \node[ConvBN] (res2a_branch1) [left = of res2a_branch2c]           {3x3 Conv\textcolor{red}{(s=1)}-BN\\ channels: 256};
    \node[plusReLU] (res2a_out) [below  =4.cm of resstage1input]               {elementwise add - ReLU};
    
    \node[draw=red, fill=gray!60, fill opacity=0.1, fit=(res2a_branch2a) (res2a_out) (res2a_branch2c)(res2a_branch1)] (stage1block1) {};
    \node[emptyAnchor, fill=none] (stage1block1text) [below right = 0cm and 0.5 cm of stage1block1.north west] {\textbf{1st block}}; 
    \node[ConvBN] (res2b_branch2a) [below right =0.5cm and -1.7cm of res2a_out]           {3x3 Conv-BN-ReLU  \\ channels: 64 (bottleneck)};
    \node[ConvBN] (res2b_branch2b) [below = of res2b_branch2a]           {3x3 Conv-BN-ReLU  \\ channels: 64 (bottleneck)};
    \node[ConvBN] (res2b_branch2c) [below = of res2b_branch2b]           {3x3 Conv-BN \\ channels: 256};
    \node[plusReLU] (res2b_out) [below  =4.cm of res2a_out]               {elementwise add - ReLU};
    
    \node[draw=red, fill=gray!60, fill opacity=0.1, fit=(res2b_branch2a) (res2b_out) (res2b_branch2c)](stage1block2) {};
    \node[emptyAnchor, fill=none] (stage1block2text) [below right = 0cm and 0.5 cm of stage1block2.north west] {\textbf{2nd block}}; 
    
    \node[draw=red, line width=2pt, fill=gray!60, opacity=0.1, minimum height=1.cm, minimum width=4.5cm] (stage1block3) [below = 0.5cm of res2b_out] {};
    \node[emptyAnchor, fill=none, text width=4cm] (stage1block3text) [below right = -0.9cm and 0.7cm of stage1block3.north west] {\textbf{3rd block} \\ Exactly the same the 2nd}; 
    \node[emptyAnchor] (resstage1bottom2) [below=of stage1block3]               {};

    \node[fill=yellow!30, fill opacity=0.2, fit=(resstage1input) (res2a_branch1) (res2b_branch2c) (stage1block3)](box) {};  
    \draw [->] (data_input) -- (conv11);
    \draw [->] (conv11) -- (conv12);
    \draw [->] (conv12) -- (pool1);
    \draw [->] (pool1) -- (resstage1);
    \draw [->] (resstage1) -- (resstage2);
    \draw [->] (resstage2) -- (resstage3);
    \draw [->] (resstage3) -- (fc1);
    \draw [->] (fc1) -- (fc2);
    \draw [->] (fc2) -- (fc3);
    \draw [->] (fc3) -- (output);
    \path [myline, dashed, shorter]{ (box.north west)  edge (resstage1.north east) edge (resstage1top2)};
    \draw [->] (resstage1top2) -- (resstage1input);
    \draw [->] (res2a_branch2a) -- (res2a_branch2b);
    \draw [->] (res2a_branch2b) -- (res2a_branch2c);
    \path[myline,->,shorter] {[out=270,in=90] (resstage1input)  edge (res2a_branch1) edge (res2a_branch2a)}
    {[out=270,in=90] (res2a_branch1)  edge (res2a_out)  (res2a_branch2c) edge (res2a_out)}
    ;
    \draw [->] (res2b_branch2a) -- (res2b_branch2b);
    \draw [->] (res2b_branch2b) -- (res2b_branch2c);
    \path[myline,->,shorter] {[out=270,in=90] (res2a_out) [bend right] edge (res2b_out)}
    {[out=270,in=90] (res2a_out)  edge (res2b_branch2a)}
    {[out=270,in=90] (res2b_branch2c) edge (res2b_out)};
    \draw [->] (res2b_out) -- (stage1block3.north);

    \path [myline, dashed, shorter] { (box.south west)  edge (resstage1.south east) edge (resstage1bottom2)};
    \draw [->] (stage1block3.south) -- (resstage1bottom2);
    
    \node[draw=blue, fit={(box.west) (box.east)}, minimum height=2cm] (explanation) [below right =-0.5cm and 3cm of data_input]           {Conv: Convolution. Stride is 1 (s=1) unless specified\\ BN: Batch Normalization\\ ReLU: Rectified Linear Unit\\ FC: Fully Connected};
\end{tikzpicture}

\caption{
 Visualization of the modified ResNet. 
  There are three stages with residual blocks, mostly like the original design, and we only show the details of the first one for simplicity. 
  In the first block within each stage, we have two Conv-BN-ReLU layers with a small number of channels (bottleneck) for better learning, and another Conv-BN layer with more channels, whose output is taken a sum with the skip connection. 
  In the following blocks the skip connection does not perform any operation.
  }   
\label{fig:NeuralNetworkArchitecture}
\end{figure*}
}

\def \tableConfusionMatricesALL{
\begin{table*}
\centering
\scalebox{0.8}{
\begin{tabular}{lllllll}
\toprule
Method &   Prediction & Has Cluster (Truth) & No Cluster (Truth) & Precision (Purity) & Recall (Completeness) & F1\\
\midrule
\midrule
   MF &  Has Cluster & 1133 & 729 & 0.61 & 0.61 & 0.61\\
       &  No Cluster & 712 & 102609 & & &\\ \\

   CNN &  Has Cluster & 1118 & 780 & 0.59 & 0.61 & 0.60\\
       &  No Cluster & 727 & 102558 & & \\ \\

   MF+CNN &  Has Cluster & 1283 & 625 & 0.67 & 0.70 & 0.68 \\
    Ensemble AND    &  No Cluster & 561 & 102713 & &\\ \\

   MF+CNN  &  Has Cluster & 1429 & 946 & 0.60 & 0.77 & 0.68\\
    Ensemble rankproduct   &  No Cluster & 416 & 102392 & &\\
\bottomrule
\end{tabular}
}
\caption{A summary of all the confusion matrices. All confusion matrices are given by selecting the threshold for each method on the validation set to maximize F1 score, and then apply the thresholds on the test set.
}
\label{tbl:ConfusionMatricesALL}
\end{table*}
}

\begin{document}

\title{DeepSZ: Identification of Sunyaev-Zel'dovich Galaxy Clusters using Deep Learning}



\author[Lin et al.]{
Z. ~Lin$^{1}$\orcidicon{0000-0001-8673-6868}\thanks{Contact e-mail: \href{mailto:zhenlin4@illinois.edu}{zhenlin4@illinois.edu}},
N. ~Huang$^{2},$
C.~Avestruz$^{3,4}$\orcidicon{0000-0001-8868-0810},
W. L. K.~Wu$^{5,6}$\orcidicon{0000-0001-5411-6920},
S.~Trivedi$^{7},$
\newauthor
\text{ }J.~Caldeira$^{8},$
B.~Nord$^{5,8,9}$\orcidicon{0000-0001-6706-8972}\\\\
${^1}$\uiuccs\\
${^2}$\berkeley\\
${^3}$\umichigan\\
${^4}$\umlctp\\
${^5}$\kicp\\
${^6}$\slac\\
${^7}$\mit\\
${^8}$\fnal\\
${^9}$\uchicago\\
}

\pubyear{2021}

\maketitle

\begin{abstract}
Sunyaev Zel'dovich (SZ) identified galaxy clusters are a key ingredient in multi-wavelength cluster cosmology. 
We present and compare three methods of cluster identification: the standard Matched Filter (MF) method in SZ cluster finding, a method using Convolutional Neural Networks (CNN), and a `combined' identifier.
We apply the methods to simulated millimeter maps for several observing frequencies for an SPT-3G-like survey and quantify differences between methods.
The MF method requires image pre-processing to remove point sources and a model for the noise, while the CNN method requires very little pre-processing of images.  Additionally, the CNN requires tuning of hyperparameters in the model and takes cutout images of the sky as input, identifying the cutout as cluster-containing or not.  
We compare differences in purity and completeness. 
The MF signal-to-noise ratio depends on both mass and redshift. Our CNN, trained for a given mass threshold, captures a different set of clusters than the MF, some of which have SNR below the MF detection threshold.
However, the CNN tends to mis-classify cutouts whose clusters are located near the edge of the cutout, which can be mitigated with staggered cutouts.
We leverage the complementarity of the two methods, combining the scores from each method for identification.
The purity and completeness of are both 0.61 for MF, and 0.59 and 0.61 for CNN.  
The combined classification method yields 0.60 and 0.77, a significant increase for completeness with a modest decrease in purity. We advocate for combined methods that increase the confidence of many lower signal-to-noise clusters.

\end{abstract}



The abundance of galaxy clusters is sensitive to cosmological parameters \citep{allen11}. Galaxy clusters have provided cosmological constraints with data from multiple wavelengths including X-ray \citep{vikhlinin09}, microwave \citep{bocquet19,planck16}, and optical \citep{costanzi19}.  
One potential systematic uncertainty in cluster-based cosmology is the selection function of observed clusters.  Clusters observed in millimeter maps have one of the better understood selection functions, providing a sample selected based on SZ signal significance, which is highly correlated with mass.

Galaxy clusters are collections of galaxies ensconced in a halo of dark matter, which provides most of the gravitational potential. Amidst the galaxies in a cluster, there exists a hot intracluster medium that emits in the X-rays (via Bremsstrahlung), and which makes them observable in the millimeter through the Sunyaev-Zel'dovich (SZ) effect. The SZ effect is an upscattering of cosmic microwave background (CMB) photons that shifts the CMB black-body spectrum along the line of sight of a galaxy cluster \citep{carlstrom02}. 
The SZ effect is independent of redshift and dependent only on the temperature of the intracluster medium, a quantity strongly correlated with cluster mass. 
An SZ-selected galaxy cluster sample therefore provides what is close to a mass-limited selection function, which is straightforward to incorporate in cosmological analyses.  
SZ-selected clusters have resulted in a number of astrophysical studies as well.  
Follow-up observations probe the physics of the intracluster medium and cluster galaxies from data in other wavelengths.  
SZ cluster follow-up include {\it Chandra} \citep{mcdonald18} or XMM-Newton \citep{bulbul19} in the X-ray and the {\it Hubble Space Telescope} \citep{strazzullo19} in the optical and {\it Spitzer} \citep{strazzullo19} or {\it Herschel} \citep{zohren19} in the infrared.  

The traditional method of identifying SZ clusters is to deploy a matched filter based method (MF) on the maps \citep{melin06,melin12}, which identifies regions in the maps that maximize the signal-to-noise ratio corresponding to the filter shape.  
The method has successfully identified cosmological samples in maps constructed with survey data from the South Pole Telescope \citep{bleem15,huang19}, Planck \citep{planckszsources16}, and the Atacama Cosmology Telescope \citep{hilton18}.
The SPT-3G camera, deployed in 2017, dramatically increases mapping speed over previous cameras. 
It is expected that there will be 5000 cluster detections at 97\% purity in the 1500 sq. deg. survey area \citep{2014SPIE.9153E..1PB}.
Next-generation experiments, like CMB-S4, will have lower noise and will likely be able to see even more objects \citep{2016arXiv161002743A}.

Convolutional Neural Networks (CNNs) are quickly becoming an essential tool for cosmology and astrophysics \citep{ntampaka19wp}.  CNNs have already been used for both CMB analyses, e.g. by \cite{caldeira19, krachmalnicoff19, hortua19}, and analyses related to galaxy clusters.
Recent applications of CNNs to galaxy clusters include mass estimations from mock X-ray images \citep{ntampaka19} and velocity dispersion distributions \citep{ho19}.
Complementary to mass estimation analyses, \citet{green19} used machine learning methods to identify physically relevant features in X-ray images that correspond to a galaxy cluster dynamical state.  
However, applications of machine learning to galaxy cluster observables in the CMB are still emerging.  

Examples of machine learning to galaxy clusters in the CMB include \citet{hurier17}, where neural networks produced filtered and cleaned maps to improve resulting cluster catalogs with lower mass thresholds and therefore higher redshifts, a proof-of-concept deep learning application to identify SZ clusters in Planck survey data \citep{2020A&A...634A..81B}, and to cluster mass estimates from CMB lensing \citep{2020arXiv200513985G}.  We emphasize that our work is the first to explicitly use deep neural networks for the identification of SZ clusters in their image space from millimeter maps in the absence of additional map filtering or cleaning steps.  In particular, we use convolutional neural networks (CNNs) to identify cluster-containing cutouts of the simulated CMB sky.  We compare our results with the standard cluster-finding method in the CMB, which has complementary performance, and explore the benefits of combining cluster-finding methods with an example implementation of such combinations and the resulting comparisons.

In this work, the CNN classification is binary, with the CNN output corresponding to a rank-ordered likelihood for a cutout of the sky to contain a cluster above a mass threshold versus not.  However,  the standard MF cluster-finding method assigns detection significance as a function of SNR, which increases with cluster mass.  As such, an apple-to-apples comparison between the two methods is admittedly artificial. We do devise a consistent way of comparing the two methods that highlights their respective strengths.  But, we note that, for a future work, a CNN regression method that predicts the mass of a cluster would more naturally lend itself to an apples-to-apples comparison with the standard MF output.

We highlight a few innovative areas of this work. We have devised a new and effective training approach for extremely unbalanced samples. We introduce a metric, the F1 score, that assesses the combined completeness and purity of the cluster sample.  The F1 score efficiently summarizes the effectiveness of the cluster finder, enabling a comparison between the CNN and MF methods.  We also use the F1 score to evaluate a combined method that incorporate both outputs from the CNN and the MF.  One can apply this approach to combine other machine-learning methods with a standard method of cluster-finding.

The paper is organized as follows.  
\mysec\ref{sec:data} describes the dataset we use to train and test the network.  
\mysec\ref{sec:method} describes both the traditional matched filter method used to detect SZ clusters and our neural network and training for our deep learning model.  
\mysec\ref{sec:results} describes our results for the network alone and the ``combined-classifier'' that incorporates results from the matched filter method.  
We discuss the implications of the results 
and summarize our paper in \mysec\ref{sec:summaryandoutlook}.
The codes we used are published on \href{https://github.com/deepskies/deepsz}{https://github.com/deepskies/deepsz}.


\section{Data}
\label{sec:data}

In this section, we discuss the origin and preparation of the data.

\subsection{Simulations}

We take simulations of the microwave sky from \citet{2010ApJ...709..920S}, which are built on top of an N-body simulation.  
Briefly, the N-body simulation used for the sky simulation has box size $L = 1000h^{-1}$Mpc, with $1024^3$ particles with particle mass $6.82\times10^{10}h^{-1}M_\odot$ and softening length $\epsilon=16.276 h^{-1}$kpc.  
These simulations provide a full-sky realization of the lensed CMB, galactic dust, point sources, and the SZ effect (both kinematic and thermal) for observing frequencies 27, 30, 39, 44, 70, 93, 100, 143, 145, 219, 225, 280, 353 GHz.

The sky simulations include the SZ signal from galaxy clusters, with halo virial masses and locations identified in the N-body simulation using a friends-of-friends (FoF) halofinder, with a linking length 0.2 of the mean interparticle spacing.
The data products we use include separate all-sky maps for each component, as well as catalogs for the locations of N-body halos, and point sources.
The halos and point sources are only unique on one octant of the sky (the other octants use various reflections of the catalogs).  We therefore restrict our search to only one octant of the sky.
We further restrict our search to the 90, 148, and 219 GHz channels, motivated by typical ground-based CMB telescopes.
In addition to the simulated sky signals, we create white noise realizations to imitate the effect of instrumental noise.
The instrument noise levels are 2.8, 2.6, and 6.6 \muka\ for 90 GHz, 148 GHz, and 219 GHz maps, respectively --- consistent with projected performance for the SPT-3G camera \citep{2014SPIE.9153E..1PB}.  All data used in our analysis can be downloaded from \url{https://lambda.gsfc.nasa.gov/toolbox/tb_sim_ov.cfm} 

For the purposes of our search, everything except the SZ signal from high-mass halos is a noise term.
On large angular scales ($\gtrsim 10$ arcmin), the maps are dominated by the CMB (because our maps are in units relative to the CMB temperature, the unlensed CMB map for each band is identical).
On arcminute scales, the maps are dominated by point sources.
Thermal sources (dusty star-forming galaxies and galactic dust) are brighter at higher frequencies, while radio sources (radio-loud galaxies, typically AGN) are brighter at lower frequencies.
The SZ signal from galaxy clusters occupies the space between point sources and the CMB, with angular scales typically between one and ten arcminutes.
In the three bands used in this work, the SZ signal is negative, and most significant at 90 GHz.
The 219 GHz channel is aligned with the null of the SZ spectrum, and has very little thermal SZ signal.
These simulations contain thermal SZ contributions from a large number of low-mass clusters, which are well below the detection threshold.
We must take these into account as an additional noise term.
The kinematic SZ signal is much smaller than any of the other noise terms.

To make these simulations more realistic, we include a noise term based on the predicted instrumental noise from the full SPT-3G survey (as noted above).
However, there are additional instrumental effects that we have ignored, which must be accounted for in real data.
First, we have used the simulations at their native resolution (approximately $0\farcm5$), without including the instrument beam.
The beams from the SPT are well-represented by Gaussians with a full-width-at-half-maximum of $\sim 1\farcm0$.
Second, the map-making process used by the SPT collaboration includes time-domain filtering to remove low frequency noise due to both the instrument and the atmosphere.
In the map domain, the filtering is approximately represented by an anistropic filter that preferentially removes long wavelength modes in the R.A. direction.
Finally, the remaining noise is not white, but increases at larger scales.
None of these effects are included in our simulations.

\figureImagesTrainingSampleExamplesPositiveNegative

\subsection{Data preparation}\label{sec:data_prep}

Data preparation for the CNN model is simpler than for the Matched Filter. 
Model training for the network {\it does not require any special pre-processing of the maps such as normalization, or point source removal}. 
For example, instead of affecting manual normalization, the usual steps in training CNNs such as convolution, batch normalization etc., implicitly process the images in a manner suitable for the prediction. On the other hand, the Matched Filter method does in fact require data preprocessing, such as point source removal, as described in Section~\ref{sec:method:matchedfilter}.  

To construct maps for the CNN, we take components and simply add them together.
First, we start with maps with just the CMB and tSZ. 
We then add instrument noise to the maps. 
We then progressively add infrared galaxies, radio sources, and galactic dust emissions to the cutouts cumulatively to facilitate investigation into which of these components have the largest effect on cluster identification.
The maps used for the MF are described in \mysec\ref{sec:method:matchedfilter}.

Our data set consists of overlapping patches (cutouts) of the mock sky.  
To prepare our data set, we cut the sky into 8 arcmin $\times$ 8 arcmin squares, with the resolution set to 32 pixels on a side.  
Cutouts are staggered with a stride of 6 arcmin.  
In other words, neighboring maps share 2 arcmin on the edge with each other.

Any given cutout likely contains a bound object somewhere along the line of sight.  
For the purposes of SZ cluster identification, we label cutouts as ``positive'' under the following conditions.  
We first consider cutouts containing clusters whose footprint in the sky is sufficiently small compared with the size of the cutout, by setting a redshift threshold.
From this catalog and cutout, our selection consists of clusters at redshift above $z\geq 0.25$ and with virial mass above M$_{\mathrm{vir}}\geq 2\times 10^{14} M_{\odot}$. 
Furthermore, we condition a cutout as ``positive'' only if the cluster position is located within the 6$\times$6 arcmin region at the center of the cutout.

Our conditions for a positive cutout ensure that most of the cluster's on-sky footprint is contained in the cutout image, thereby reducing potential edge effects.  
Since the distance between the centers of our cutouts is 6 arcmin, each cluster in our mass and redshift range is contained in exactly one ``positive'' cutout. 

Given our specific choice of mass and redshift threshold, only around 1\% of the cluster-containing cutouts contained more than one cluster that fit our threshold. A potential direction one could take is to iteratively train models where the thresholds for cluster-containing cutouts change.  However, this would result in multiple clusters per cutout, complicating comparisons with the MF results.  We therefore choose the mass and redshift thresholds specified for the label ``cluster-containing'' to simplify the performance comparison with the MF.  To enable a straightforward treatment of purity, completeness, and F1 score, we assume a bijective correspondence between a true positive prediction in the data set and an actual halo from the original simulations.

With this procedure, our dataset contains 808,201 total cutouts. Out of these, 14,989 are labeled as positive.  
Note that the positive samples form a small percentage of the full dataset (less than two percent).  
In machine learning, this is referred to as an \emph{unbalanced} dataset since the class of ``Negative'' labels has many more 
objects than the class of ``Positive'' labels.  
Unbalanced datasets pose an additional challenge when training neural networks.  
We discuss this challenge and our approach in a later section (\mysec\ref{sec:matching_to_truth}), where we have identified a metric that is not sensitive to the class imbalance.

The next stage of data preparation is to split our dataset into training, test, and validation sets. 
We group cutouts by their position in the sky.  
The \emph{training} set is comprised of cutouts with center right ascension (RA) coordinate above $0.2\times90$ deg, the \emph{test} set comprised of cutouts with center RA coordinate below $0.13\times90$ deg, and the remaining cutouts grouped into the \emph{validation} set.
To avoid training a network on the periphery of a cluster, in the training set we remove the negative cutouts that are adjacent to a positive one. 
We do not do this on the validation or test set.
The final numbers of cutouts in the training, validation and test sets are 601,400, 105,183, and 56,637 respectively.

\myfig\ref{fig:ImagesTrainingSampleExamplesPositiveNegative} shows two sets of example maps at 90 GHz, 148 GHz and 219 GHz.  The top half of this figure shows cutout maps with a galaxy cluster, where the true position of the cluster is marked with a red ``X''.  The lower panel of this figure shows example cutout maps that do not contain a galaxy cluster.  From left to right, we add additional sources of noise.  The addition of IR and radio galaxies increases the visual inhomogeneity of the image.

We next discuss the impact of high flux sources. 
Figure~\ref{fig:ImagesTrainingSampleExamplesPositiveNegativeInfrared} shows an example cluster-containing cutout from our sample that happens to also host a high flux radio galaxy.   The column on the left is the image with only the CMB+tSZ components simulated, similar to the left-most column of Figure~\ref{fig:ImagesTrainingSampleExamplesPositiveNegative}.  The column on the right has all other components added, similar to the right-most column of Figure~\ref{fig:ImagesTrainingSampleExamplesPositiveNegative}.  In the absence of the high flux radio galaxy (left column), the tSZ signal from the galaxy cluster is visually identifiable in the 90 GHz and 148 GHz frequencies and the colormap for these images spans a few hundred $\mu$K.  In contrast, the images that include the high flux radio galaxy (right column) span the thousands of $\mu$K, and the tSZ signal from the galaxy cluster is barely identifiable in the only the 90 GHz frequency with the same linear scaling.  The high flux galaxy is the main visible feature in these images, seen as two very bright pixels near the center of the cluster.

The MF method requires a preprocessing step of point-source removal in order to identify the high signal-to-noise co-located pixels of the tSZ signal from galaxy clusters.  Many classical machine learning methods also require some sort of preprocessing, such as rescaling and normalization, to optimally perform.  We emphasize that the CNN method we use does not require any such manual preprocessing and takes as input, pixel data from images like the high flux galaxy containing cluster.  This cutout is an example of a ``True Positive'' identified cluster by our method.

\section{Methods in SZ Cluster-finding}
\label{sec:method}
\figureImagesTrainingSampleExamplesPositiveNegativeInfrared %
We next describe distinct methods of SZ galaxy cluster detection --- first the canonical method, Matched Filter (MF), and then an alternative method via Deep Convolutional Neural Networks (CNNs). 
We are developing the methods in simulations, and therefore know the ground truth of all the clusters, including, namely, the properties of their underlying dark matter haloes. 
To assess cluster-finding efficacy, we must match clusters to haloes, which we describe after the detection methods.
We then note the differences and similarities of the methods in principle, in practical implementation, and compare their efficacies.

\subsection{Matched Filter (MF)}
\label{sec:method:matchedfilter}

We applied a matched-filter-based method to the simulated microwave maps. 
The matched-filter is a standard method to identify galaxy clusters from their SZ signature, providing a baseline to compare our deep learning model.
It uses the spectral and spatial characteristics of the SZ signal from galaxy clusters to differentiate them from noise.
This method was developed in \citet{melin06}.
We apply the matched filter method as applied to data collected with the South Pole Telescope (e.g. \citealt{vanderlinde10} (hereafter V10), \citealt{bleem15}, and \citealt{huang19} (hereafter H19)).
In this section we provide a descriptive overview of the method and leave more mathematical and detailed treatments to the above references.

A matched filter is a Fourier-domain filter which optimally filters for a given input profile, and a given set of noise power spectra.
The input profile is weighted by the inverse of the noise power spectra.
This weighting scheme provides an optimal filter under two conditions: the noise terms are Gaussian and stationary (that is, the noise power spectra do not vary over the map).
In order to maximize signal-to-noise on SZ clusters, our matched filter also combines the maps across frequency bands.
For each band, we model the following noise terms:
\begin{description}
    \item[{\bf CMB:}] 
        The CMB contributes noise primarily at large angular scales.  
        We calculate the CMB power spectrum using the Code for Anisotropies in the Microwave Background (CAMB; \citealt{lewis00}) with the best fit \lcdm\ parameters from WMAP7+SPT \citep{komatsu11, keisler11}.
        We also include a term for the kinematic SZ, based on \citet{shirokoff11}.
        These terms have the same amplitude in each band.
        The simulations used in this work generate the CMB realization based on WMAP5 \lcdm\ parameters.
    \item[{\bf Point Sources:}]
        The brightest point sources contribute noise on the smallest scales, but the background of unresolved sources contributes to a much broader range of scales.
        We model the combination of two source populations: one from dusty galaxies, which are brighter at higher observing frequencies; and one from radio-loud sources, which are dimmer at higher observing frequencies.
        We assume that the spatial power spectrum of the combined source population is flat in $C_{\ell}$-space.
        Each frequency is normalized such that $D_{\ell} = \ell \left(\ell + 1\right) / \left(2 \pi\right) C_\ell = $ (2.7, 8.8, 71.) \ukcmb$^2$ at (90, 148, 219) GHz and $\ell = 3000$.
        These values were chosen to match the amplitudes of the point source power spectra from the simulated maps used in this work.
    \item[{\bf SZ Background:}]
        There is a contribution to the noise from dim, undetected SZ sources.
        We model this as flat in $D_\ell$-space, with $D_{3000} = 3.6$ \ukcmb$^2$ at 90 GHz, and the remaining bands scaled from this value using the non-relativistic form of the SZ frequency spectrum.
    \item[{\bf Instrumental Noise:}]
        Instrumental noise is also flat in $C_\ell$-space, with amplitudes given in \mysec\ref{sec:data}.
\end{description}
Point sources are the primary source of non-Gaussian non-local noise.
In previous applications of this method, point sources with high signal-to-noise ratio were masked to avoid the false detections they cause.
Due to the low noise levels assumed, the threshold for point source masking is much lower, which would lead to masking a significant fraction of the map area.
Instead, we have subtracted point sources that are brighter than 2 mJy before filtering the maps.
This is consistent with the projections for finding clusters using the SPT-3G receiver.

To find clusters, we filter the maps using a projected $\beta$-model \citep{cavaliere76} with $\beta = 1$.
V10 explored more complex models, but found no increase in the efficacy of the matched filter.
The $\beta$-model has one free parameter, which sets the angular scale of the profile.
We create 12 different filters to account for clusters of different angular sizes.

Our filtering procedure produces a set of 12 maps (one for each profile) in signal to noise units.
We run a peak-finding algorithm which groups connected pixels above a given SN threshold.
Each group's position is calculated by taking the SN-weighted mean of the pixel positions.
Finally, for groups with detections in multiple output maps, we take the group with the highest SN.
These detections make up the final cluster candidate list.
Each candidate has a location, SN (which is used as a proxy for mass; see e.g. H19), and the profile which maximized its SN.

\subsection{Deep Convolutional Neural Networks (\cnnabbreviation)}
\label{sec:method:deepneuralnetwork}

The machine learning model we have chosen for this work is a deep convolutional neural network, which is known to perform well in image classification tasks. 
Compared with the matched filter method, neural-network-based methods require less image pre-processing, and are very flexible in that the same architecture can be used in different tasks with little modification. 
For example, model structures developed in classification problems are often used for regression problems. 
In fact, even the weights trained on one task can often help in the training of a different task.
With this in mind, we will use a well-known architecture as a starting point for our network structure.

\subsubsection{Network Architecture: ResNet-50}

Our network design is based on ResNet50 \citep{he16}, which is a powerful and popular deep convolutional neural network architecture \citep{lecun98}. 

A typical \textit{neural network} takes in a multi-dimensional array (e.g. a RGB image as a 3D array), and outputs a scalar or array, depending on the use case. 
In our binary classification case, the output is a real number denoting the ``score'', which should be a rank-preserving function of the probability that a cutout contains a target.
A \textit{deep neural network} usually consists of several layers, and in the simplest case, each layer has several \textit{neurons}, each of which computes a linear combination of its inputs, and then passes the value through a non-linear function.
This output then forms the input of the next layer.

A \textit{deep convolutional neural network} (DCNN or CNN) is characterized by the presence of convolutional layers, possibly among others.
A convolutional layer takes in an image or a batch of images as the input. 
A layer contains small (usually several pixels by several pixels) learnable filters, and convolves each filter with the input using a sliding window, usually with a stride, to get a feature map.
This feature map output is then passed through an element-wise non-linear function.
Beyond the first layer, the input are feature maps generated by previous layers instead of images.
A convolutional layer has a few notable distinctions compared with a fully connected layer: It reduces the number of parameters (parameter sharing), and, partly because of this, its output is less sensitive to translation in the input (\textit{translation invariant}), which proves to be a very desirable property in applications like image classification.
After a convolutional layer, a typical CNN will have a pooling layer that takes the maximum activation in a small region (usually 2x2 pixels). 
Strides in the convolutional layer and pooling layer are two ways to down-sample the feature maps, and by repeating these steps, the neurons in the deeper layers can ``see'' a larger and larger region of the original image.

If layers are just stacked to build a very deep neural network with no other modifications, we often see an degradation in accuracy, since the parameters become very difficult to optimize as the network becomes deeper.
Note that while the number of parameters grows with the number of layers, the phenomenon referred to here is manifested in a higher \textit{training} error, so this is not related to overfitting.
A \textit{Residual Network} (ResNet) is a DCNN that aims to mitigate this issue.
A Residual Network has several residual blocks, each consisting of a few convolutional layers and a shortcut connection from the first layer to the last within the block.
The idea is that the shortcut connection behaves like an identity function, and the layers inside a residual block will only need to add what has not been learned by layers prior to the block: the ``residual'' of this identity map. 
Shortcut connections then allow us to make models deeper without degrading performance.
This is because simply adding identity layers to a model (that is, if the additional layers that are skipped add nothing new) will not increase the \textit{training} error.

ResNet50, proposed in the original ResNet paper, is a popular ResNet model used in many different image classification tasks.
It has 50 convolutional or fully connected layers in total, grouped into several residual blocks.
The depth of ResNet50 was optimized to train on relatively big cutouts.
However, to cater to the small image sizes in our dataset, we effect the following changes: 
\begin{itemize}
    \item[1.] We remove the first convolution layer (kernel size 7, stride 2, padding 3), and replace it with 2 smaller convolution layers with kernel size 3 and padding 1. 
    The first convolution layer has a stride of 1 whereas the second has a stride of 2. 
    Like the original ResNet50, we have batch normalization and ReLU after each convolution layer, and we also have a max pooling layer of stride 2 after the second convolution layer. 
    \item[2.] Instead of having a 5th stage, we put 2 fully connected layers of size 256 on top of stage 4's output, and put a prediction and softmax for 2 classes afterwards.
\end{itemize}
\noindent
\myfig\ref{fig:NeuralNetworkArchitecture} shows the structure of the network visually. 
\figureNeuralNetworkArchitecture

\subsubsection{Training Details}
As mentioned earlier, our data is highly unbalanced. The ratio of negative cutouts to positive cutouts is over a factor of 50.
To address this challenge, we experiment with several training strategies. 
We first manually assign different weights to positive and negative samples to give comparable importance to the positive and negative subsamples.  Similarly, we oversample the positive samples to establish an effective negative-to-positive ratio closer to 1. 
However, these strategies result in a trade-off. 
If we set the effective ratio low (closer to a balanced sample), then the network carries a bias unrepresentative of the real sample. 
On the other hand, if the effective ratio is too high, then the network is very difficult to train.  This difficulty comes from the fact that a blind guess of all negatives can already produce a good accuracy and loss.
Moreover, iterating the process over different ratios is very time-consuming. To solve this problem, we devise the following strategy:
\begin{enumerate}
    \item Let us denote $r$ as the effective ratio of the number of negative cutouts to the number of positive cutouts. 
    $S_{r}$ is a sample where the positive cutouts are over-sampled such that the negative-to-positive effective ratio is $r$. 
    One can think of this as a data-loader of the same training set that gives positive cutouts a higher probability of being sampled.
    \item We start by training the model on $S_1$, a sample where the there is an equal probability of drawing a negative or a positive cutout. 
    We train the model with a batch size of 128, using cross-entropy loss.
    We evaluate the loss on the over-sampled validation set every 100 batches and continue training until the loss on the validation set converges (i.e. does not decrease for 1000 batches). 
    We consider this as an epoch. 
    \textit{Note that each epoch can have different sizes.} 
    As an example, the first epoch took 2400 batches to complete.
    \item In each epoch, we use stochastic gradient descent for training with an L2 weight decay rate set to 0.003. 
    We set the learning rate to $5 \times 10^{-3}$, with a linear warm-up schedule in the first 500 batches in each epoch (linearly increasing from  $5/3 \times 10^{-3}$ to $5 \times 10^{-3}$). 
    After 1000 batches, we decrease the learning rate to $5 \times 10^{-4}$. 
    \item At the end of each epoch, we save the best model weight so far, feed the original validation set (no over-sampling involved) to this model and save the results. 
    We then increase $r$ by one and continue training starting with this model weight.
    \item After training on $S_{20}$, we go back and pick the model weights that best perform on the entire validation set. 
    In this experiment, the best performing model that we select is from $S_{16}$.
\end{enumerate}

We can consider this to be a dynamic stratified sampling.
The entire training process, including evaluation of the test and validation sets at the end of each ratio, took approximately 8 hours on a GTX 1060 GPU. 
We first show some of the training and validation cross-entropy loss statistics during the training process in the top panel of \myfig\ref{fig:NeuralNetworkTrainingHistory}. 
We also overlay the ratio at each iteration in the plot.
In the bottom panel of \myfig\ref{fig:NeuralNetworkTrainingHistory} we show the training and validation accuracy adjusted for the ratio, along with the ratio at each iteration.
Unless specified, all numbers reported are for cutouts with all noise components. We perform a breakdown of these results in Section~\mysec\ref{sec:results}. 
Finally, we perform the classical Platt Scaling calibration method for binary classification, (\cite{Platt99probabilisticoutputs}), on the final output, so that we can interpret the output as probabilities.
This is equivalent to running a logistic regression using output on the validation set.
All CNN outputs in this paper are the calibrated numbers.

\subsection{Matching Cluster Detections to Haloes}\label{sec:matching_to_truth}
\label{sec:matching}    

The MF assigns detections to specific coordinates, while the CNN detects whether or not each cutout contains a cluster.
To compare two methods, we need to match MF detections to cutouts.

MF-identified clusters whose centers lie near the edge of a cutout may have an identified center just inside the cutout, but a true center outside the cutout region. 
In these cases, the cutout containing the MF identified cluster position would not be a ``cluster-containing'' cutout, despite the actual correspondence. 
These cases would lead to a comparison that understates the MF performance.
To ensure correspondence and to make a fair comparison between methods, we perform the following:
\begin{enumerate}
    \item We match MF detections to the largest cluster within a 1-arcmin radius of the detection. 
    The radius was chosen to maximize MF's performance.
    \item If the detection corresponds to a real cluster, we match the detection to the cutout corresponding to the true cluster center.  If the detection does not correspond to a real cluster, we match the detection to the cutout containing the MF identified center.
    \item We assign the corresponding signal-to-noise-ratio (SNR) of the MF detection to the cutout to which the detection is assigned in the step above.
    If multiple detections map to the same cutout, the largest SNR is assigned to that cutout.
\end{enumerate}


\section{Results}
\label{sec:results}
\figureNeuralNetworkTrainingHistory
Below, we present results of both methods applied to the mock data set with all noise components. 
Table \ref{tbl:ConfusionMatricesALL}\tableConfusionMatricesALL summarizes all the confusion matrices after appropriate threshold selection. These results use thresholds that maximize the F1 score of the methods applied to the validation set (see \ref{sec:f1} for the detailed procedures). Solely focusing on purity or completeness as a metric can be misleading; the F1 score offers a fairer platform for comparison. We see that the F1 score performance of MF and CNN are comparable.  Additionally, the Ensemble methods can significantly improve the performance measured by any of the metrics.  The Ensemble methods either simultaneously improve both purity and completeness (AND ensemble), or greatly improve one without sacrificing the other (PROD ensemble).

\subsection{Ambiguity in Definition of True Positives}\label{sec:amb_tp}

In this subsection, we describe one possible systematic in our comparison of the two methods. To assign a performance metric to a cluster detection method, we need a mass threshold for a ``true cluster''.  There are otherwise multiple halos in any given patch of the simulated sky, some of which sit above that threshold and some of which sit below. Due to hierarchical structure formation, there are more objects below a mass cut than above.  An artifact of a cutoff is that there will likely be some confusion near the selected threshold. 

In particular, the matched filter does not assume a clean cutoff in mass when identifying a cluster, but rather identifies a cluster according to the SNR.  The SNR produced by the matched filter scales with cluster mass (and weakly with redshift, see \citet{bleem15}, and H19) with a $\sim$20\% lognormal scatter in the scaling relation (see e.g. \citealt{bocquet19}).  To briefly describe the relationship, at fixed mass, the temperature of the cluster gas increases with redshift therefore increasing the tSZ as well.  And, at larger angular sizes, the cluster size becomes comparable to the smallest CMB fluctuations.  The MF downweights such modes, ultimately contributing to a redshift dependence in the MF performance.  
We included a redshift threshold of 0.25 (see Section \ref{sec:data_prep}) to somewhat address such issues. 

We can conjecture that the CNN will also get confused by an object whose mass is near the threshold.  We assume a mass threshold when labeling a cutout as ``cluster-containing''.  But some halos below the mass threshold may have a higher MF SNR or CNN score than halos above the detection threshold.  This ambiguity affects the precision of both methods.
Nonetheless, we fix the threshold of each method in our comparison in an effort to fairly assess each.


\subsection{F1 Score}\label{sec:f1}

In Section~\ref{sec:amb_tp}, we discussed how the ambiguity in defining true positives in the sample might affect the metrics for model performance.  Another aspect of performance metric comparisons is assessing the trade-off that occurs when we vary the classification threshold; we can increase the number of true positives, improving the completeness of a sample, at the cost of increasing the number of false positives, worsening the purity of a sample.  The cost-benefit comparison becomes more complex in an imbalanced dataset, where there is a large difference between the number of true positives and the number of true negatives.  This is the case for the number of cluster containing cutouts; galaxy clusters are relatively rare objects in the overall footprint of the sky.

The F1 score is a common choice of performance metric in problems with imbalanced samples.  If a classifier blindly predicted everything to be positive (or negative) and there were many more true positives (negatives) in a sample, we would see a misleadingly high accuracy.  The F1 score accounts for the imbalance, defined as the harmonic mean of precision and recall: 
\begin{equation}
    \text{F1} = \frac{2}{\text{precision}^{-1} + \text{recall}^{-1}} = \frac{2\text{TP}}{2\text{TP}+\text{FN}+\text{FP}},
\end{equation}
In our particular use case, a classifier could identify all cutouts as non-cluster-containing and achieve a high accuracy.  But the F1 score would equal zero, indicating the lack of information provided by this classification scheme.  A high F1 score indicates a classifier with high precision while detecting as many clusters as possible despite the imbalance between the two classes in our dataset.  We therefore emphasize the F1 score in subsequent discussions of model performance.

Note we advocate for F1 as a useful metric, regardless of the method classifier.  We can use the F1 score to determine an optimal threshold for positive predictions for a given classifier.  We do that by looking at how F1 varies as a function of the threshold for positive identification of that classifier.  The maximum F1 score corresponds to a threshold that maximizes positively identified clusters while minimizing false positives, useful for any cluster analysis that requires further follow-up observations.  The peak F1 score also provides a single metric to compare any classifier.

\figureFoneScoreComparison

\myfig\ref{fig:FoneScoreComparison} shows the F1 score on the validation set of MF and CNN as a function of the threshold for positive identification.  The MF has a threshold at some SNR, and the corresponding F1 score is shown with the purple line. The CNN has thresholds of probability output with the corresponding F1 score shown in the black line.  The shaded purple and black histograms respectively show the number of objects within a given SNR bin and probability bin.  

From this figure, we see that the peak F1 score of MF applied to the validation set is 0.60, which corresponds to a SNR threshold of 13.4, the 98.2\% rank percentile of the MF SNR of all cutouts.  That same SNR threshold results in an F1 score of 0.62 when applied to the MF SNR of the test set.  
We can also determine that the peak F1 score of CNN applied to the validation set is 0.60, corresponding to a probability threshold of 0.367, the 98.0\% rank percentile of the CNN scores of all cutouts.  
That same probability threshold results in an F1 score of 0.60 on the test set.  Using the F1 score to compare the two methods, we see that they do comparably well. We summarize the metrics in Table~\ref{tbl:ConfusionMatricesALL}.

Note, if we consider true-positives with lower cluster mass thresholds, the F1 score of the MF as a function of SNR would shift.
At fixed SNR, as we decrease the mass threshold, purity (precision) will increase, but completeness (recall) will decrease.
As a result, the direction of the shift of F1 at each SNR value will be determined jointly by these two effects.  But, the peak of the F1 score curve, i.e. the optimal SNR threshold, will systematically shift to lower SNR values since true clusters include lower mass objects.  However, our analysis does not include these results, since we want to compare the MF results with a given CNN model that was trained at a given mass and redshift threshold.  It would be computationally expensive to iteratively train more CNN models to compare with the MF performance at different mass thresholds.  Furthermore, we will have multiple clusters per cutout, complicating our comparison metrics (see discussion in Sec~\ref{sec:data_prep}.)

The F1 score also provides a means to combining the two methods for cluster identification.  We evaluated two ways to combine the predictions, an \textbf{EnsemblePROD} method and an \textbf{EnsembleAND} method described below. 

\begin{itemize}
   
    \item \textbf{EnsemblePROD} method:  Here, we form one single score by multiplying the rank-percentile of CNN and MF scores, similar to an interaction term in regression tasks. Suppose we have $N$ cutouts, and one CNN score $p_i^{CNN}$ and one MF SNR $s_i^{MF}$ for each cutout $i \in [N]$. The CNN rank percentile is defined as $q_i^{CNN} := \frac{1}{N}\sum_{j\in[N]} \mathbbm{1}[p_j^{CNN} \leq p_i^{CNN}]$, and the MF rank percentile is defined similarly - just the percentile its MF SNR falls in the sample.
    The combined rank percentile is simply their product: $q_i^{Ensemble}:=q_i^{CNN}q_i^{MF}$.
    
    \item \textbf{EnsembleAND} method: Here, we classify a cutout basing on a logical AND condition basing on CNN and MF scores. Unlike previous methods, this method requires 2 thresholds (one for CNN and one for MF), and classify a cutout as positive if both thresholds are met.

\end{itemize}

We evaluate the metrics for the combined classifiers using the same metrics and procedures as we did for the individual methods. We choose the threshold on validation set and evaluate the performance on the test set. 

\myfig\ref{fig:FOneScoreEnsemblePROD} shows the results of the F1 score curve for \textbf{EnsemblePROD} as a function of the rank product, defined above.  The peak of the curve corresponds to the rank product that results in the maximum F1 score.  We summarize the metrics in \mytbl\ref{tbl:ConfusionMatricesALL}. Recall that the thresholds picked on the validation set for the individual methods are 98.2\% for MF and 98.0\% for CNN.  The new rank-product threshold, 93.8\%, is at the 97.7\%-percentile of all rank-products, which is very close to the two standalone threshold ranks. In other words, \textbf{EnsemblePROD} gives a similar number of positive predictions (compared with MF and CNN), but the overall F1 score, as shown in Table \ref{tbl:ConfusionMatricesALL}, is greatly improved.
This suggests that \textbf{EnsemblePROD} successfully incorporates information from both methods.
    
\myfig\ref{fig:FOneScoreEnsembleAND} shows the results of the F1 score for \textbf{EnsembleAND}, color-coding the F1 score in the space of the MF and CNN thresholds for positive identification.  We summarize the metrics in \mytbl\ref{tbl:ConfusionMatricesALL}. As expected, the two thresholds are significantly lower than the counterparts in the standalone MF or CNN methods, which is a direct result of the AND logic we apply here. This however suggests that CNN predicts certain low-MF-SNR cutouts with higher probabilities and the MF measured a higher SNR for certain low CNN probability cutouts; the two methods complement one another.
    
\figureFOneScoreEnsembleMethods
We now compare the performance of the ensemble methods with the the standalone on the test set.  The F1 score allows for an overall comparison along a single dimension.  Both ensemble methods, with F1 scores of 0.69 and 0.68 respectively, lead to significant improvement over standalone MF or CNN method, with F1 scores of 0.61 and 0.60 respectively. This illustrates the strength in combining methods, particularly with the use of F1 scores.  In Section~\ref{sec:method_comp}, we investigate how the two methods complement one another.

\subsection{Comparison of Performance}\label{sec:method_comp}

\figureImagesPredictionExamples
\figureNeuralNetworkPerformanceMetrics
We compare the CNN and MF performance using thresholds that maximize the respective F1 score of each method.  With this, we arrive at the precision and recall values provided in Table~\ref{tbl:ConfusionMatricesALL}.  The first two rows of the table show the standalone performance of each the MF and CNN.  The precision of the CNN is slightly worse than the MF, and the recall slightly improved. 
We want to emphasize again that only looking at either precision or recall here would be misleading, as our evaluation metric is mainly F1 score. 
For example, it's true that the \textbf{EnsemblePROD} method does not improve precision, but that's because it trades (on the validation set) precision for a much higher recall to improve the F1 score.

Given the differences inherent to each method, we further explore if the two methods are complementary to one another, preferentially selecting (or missing) clusters with particular attributes. 
To further investigate, we looked at two sets of clusters in the extremes of the CNN and MF performance:

\begin{enumerate}    
    \item \textbf{Clusters with high MF SNRs but low CNN scores:}  The left panel of \myfig\ref{fig:split2_CNNfailures} shows $R_{\rm vir}$ as a function of cluster redshift, color coded by $t_{\rm SZ}$.  We show the clusters whose CNN score is below $0.5$ and MF SNR above $15.4$.  The right panel shows the position of each cluster within the cutouts.  These clusters are very close to the edge of the center of the cutout, which is a $6\times 6$ arcmin region, appearing as ``almost'' negative samples. To investigate this edge effect further, we randomly re-generated cutouts with clusters with different distances to the center and evaluated the CNN on these new cutouts.  We discuss this ``edge effect'' in Section~\ref{sec:edge_cnn}.
    \item \textbf{Clusters with high CNN scores but low MF SNRs:} The left panel of \myfig\ref{fig:split2_MFfailures} shows $R_{\rm vir}$ as a function of cluster redshift, color coded by $t_{\rm SZ}$.  We show the 73 clusters whose CNN score is above $0.8$ and $<11.3$ MF SNRs. 
    The right panel shows the position of each cluster within the cutouts.
    Compared with clusters with low CNN scores, these are closer to the cutout center on average.  
\end{enumerate}

\subsection{CNN Specific Considerations}
\subsubsection{Effects of noise components on CNN performance}

Given the slightly worse overall performance of the CNN, we examine noise components as a potential cause to the lower performance. 
\myfig\ref{fig:NeuralNetworkPerformanceMetrics} shows the performance of the standalone CNN with different levels of noise (different components), quantified with a ROC (receiver operating characteristic) curve. 
With each added noise component, the area under the corresponding ROC curve (AUC) decreases, as one might expect. 
However, the difference is not drastic varying only from AUC=0.99 to 0.98. 
The resulting \textbf{test} F1  we measure by using validation thresholds with each added level of noise is 0.64 for cmb+tsz, 0.63 for +ksz, 0.61 for +IR galaxies, 0.6 for +radio galaxies, and 0.61 for +galactic dust. 
In both cases, infrared galaxies and radio galaxies seem to affect the performance of CNN the most.  We conclude that the CNN is relatively robust to the noise components in the microwave sky.

\subsubsection{Effects of edge location on CNN performance}\label{sec:edge_cnn}

As described in Section~\ref{sec:method_comp}, we found that many of the low-CNN-probability and high-MF-SNR cutouts have their cluster very close to the edge. 
We re-generated cutouts with the same cluster at varying distance from the center, and apply the same trained network on these shifted cutouts.
\myfig\ref{fig:split2_pred_over_dist} shows the positions of these clusters in the new cutouts.
Interestingly, we did find a very obvious negative correlation between the prediction score and the clusters' distance from the center of the cutouts, as shown in \myfig\ref{fig:split2_pred_over_dist}. 
This suggests potential room of improvement with some different ways to apply the network.
A potential algorithm to maximize the performance of the CNN could be to simply overlap cutouts by 1/4 the width of a cutout.  This way,  each cluster is within the middle two quarters of at least four cutouts. While several cluster containing cutouts would suffer from the edge effect, each cluster would be positively identified in some subset of the cutouts. 

\figureFailuresAndShifted

\subsection{Completeness comparison}
We now compare the cluster identification methods showing the {\it completeness} of each method, i.e. their performance as a function of cluster properties.  

\figureCompletenessMvir
Figure~\ref{fig:CompletenessMvir} shows the completeness, also known as recall, of each method as a function of cluster mass.  This is the fraction of true positives identified by a given method out of the total number of cluster images evaluated.  Each line color corresponds to a different identification method:  black, purple, and blue respectively correspond to CNN, MF, and EnsemblePROD.  The solid lines show the completeness for the entire sample, and the dashed lines for clusters above $z>0.25$.  Recall, we impose both a mass cut of $M_{\rm halo}\geq 2\times 10^{14}M_\odot$ and a redshift cut of $z>0.25$ for cutouts labeled as ``cluster-containing'' in our training set for the CNN classifier.  We mark this threshold with a red vertical line.  For comparison, the blue histogram illustrates the underlying halo mass function of the simulated galaxy cluster sample that we used to produce the cutouts, plotted as ``Counts of Objects'' (right y-axis labels) as a function of the virial mass bin.  To guide the eye for comparison, the bottom panel shows the ratio of the recall of CNN and MF with the recall of EnsemblePROD with the same redshift selection.      

In the range of $1.5\lesssim \frac{ M_{\rm halo}}{10^{14}M_\odot}\lesssim3.5$ 
the CNN has a higher recall.  At smaller or larger masses, the MF has a higher recall.  However, across the mass range, the EnsemblePROD has a systematically higher completeness compared with either the MF or the CNN.  The overall improvement indicates the complementarity of the two methods. We note that the decrease of the CNN recall (black curves) at the highest masses is due to the decreased sample size for training.  Very few of the highest mass objects are at $z>0.25$, meaning that many of the high mass objects were also not labeled as ``cluster-containing''.  

\figureCompletenessRedshift
Figure~\ref{fig:CompletenessRedshift} shows the corresponding selection as a function of redshift.  Here, we only show the completeness curves of each method for clusters that are above our mass cut for ``true'', $M_{\rm halo}\leq 2\times10^{14}M_\odot$.  The red vertical line marks the redshift threshold for cutouts labeled ``cluster-containing''. 

For clusters below $z\lesssim0.9$, the EnsemblePROD outperforms both the CNN and the MF in completeness.  At higher redshifts, the completeness of the sample selected by the MF is comparable to that of the sample selected by EnsemblePROD, both exceeding the completeness of the CNN, which remains relatively constant until $z\sim1.5$. 
The EnsemblePROD has a larger F1 value with similar completeness at high redshifts as the MF, thereby illustrating that information from the CNN helps to maintain purity in cluster identification.  In fact, for both lower mass and lower redshift galaxy clusters, the completeness of the sample improved when we combine the two methods.  This improvement suggests that the MF and the CNN are picking up complementary features in our galaxy cluster sample.   

To better assess the complementarity, we visualize the performance of each method in Figure~\ref{fig:split2_selfunc_both}.  Here, we show the distribution of the virial mass as a function of three cluster parameters in our sample.  From top to bottom, we show the virial mass as a function of redshift, virial radius, and angular size of the cluster.  From left to right, we show the percent of true positives identified by the CNN, EnsemblePROD, and MF, where we color code the parameter space by the true positive percentage.  Darker shades of green correspond to higher true positive rates in that region of parameter space. Each pixel in the color coded parameter space corresponds to at least 3 clusters from our sample.  For reference, the horizontal dashed line indicates the mass threshold for clusters in cutouts that we labeled as ``cluster-containing''.   

If we look at the performance of the methods in the Virial Mass vs. Redshift space, we can see that the CNN positively identifies more of the clusters at low redshifts whose masses lie just above the mass cut than the MF identifies (i.e. 
$2\times10^{14}M_{\odot}\lesssim M_{\rm vir}\lesssim 3\times10^{14}M_{\odot}$ 
and $z\lesssim0.7$).  On the other hand, the MF positively identifies more of the clusters at higher redshifts whose masses lie just above the mass cut (i.e. 
$2\times10^{14}M_{\odot}\lesssim M_{\rm vir}\lesssim3\times10^{14}M_{\odot}$  and $z\gtrsim 1$).  As a result, the EnsemblePROD has more positive detections of clusters above the mass cut across the redshift range.  

We can do a similar comparison in the Virial Mass vs. Virial Radius space.  Here, we see that the MF preferentially picks out the highest mass halos at fixed Virial Radius, missing some of the clusters near the mass threshold that are more spatially extended.  The CNN manages to identify more of these ``missed'' clusters, with identified clusters that appear to be more complete with a limiting mass.  Again, we see how the complementarity manifests itself in the EnsemblePROD, which identifies clusters missed by either method on its own.

The last row of this figure, showing the percent of positively identified clusters in the space of Virial mass vs. Angular Size, summarizes the effects from the first two rows.  Clusters with large angular size are lower redshift clusters with larger virial radii. These are ``missed'' clusters that the CNN identifies more than the MF, leading to a better performance for the EnsemblePROD.   

\section{Summary and Discussions} \label{sec:summaryandoutlook}

In this paper, we compare the performance of a matched filter (MF) method and a convolutional neural network (CNN) in the task of identifying galaxy clusters in mock millimeter maps of the cosmic microwave background.  For the neural network, we use a modified version of ResNet, a architecture used in popular image classification.  We use simulated microwave maps at 90, 148, and 219~GHz channels with added observational components to train and test our CNN.  We also use the F1 score  (see section~\ref{sec:f1}), a quantity that accounts for both precision (purity) and recall (completeness), to compare method performance and to define an identification procedure that combines both the MF and CNN.  We find the following:

\begin{itemize}
    \item At the selected redshift and mass thresholds, the CNN does comparably to the MF (see Table~\ref{tbl:ConfusionMatricesALL}).  The precision (purity) of the CNN is slightly lower, but the the recall (completeness) slightly higher.
    \item The CNN achieved comparable performance in the absence of standard image pre-processing, e.g. normalization, point source subtraction, etc. 
    \item A cluster identification procedure that combines both the MF and CNN scores significantly improves performance (e.g. see Figures~\ref{fig:CompletenessMvir}, \ref{fig:CompletenessRedshift} and \ref{fig:split2_selfunc_both}), indicating complementarity between the methods.
    \item We note that the cutout nature of the train/test/validation dataset for the CNN impacts the model performance; clusters further from the cutout center are more difficult for the CNN to identify (see section~\ref{sec:edge_cnn}).  An algorithm that minimizes the distance between the cluster center and the cutout center would further improve the CNN performance. 
\end{itemize}

We note that some cutouts contributing to false positive identification are cutouts that contain clusters just below the mass threshold of $M_{vir}=2\times10^{14}$ that we label as ``cluster-containing''.
The cluster-finding task differs from most other standard classification applications in that galaxy clusters may be defined on a continuum; 
the separation between galaxy clusters and galaxy groups is largely definition-based.  
Admittedly, galaxy clusters would be ideal objects to apply regression methods that predict continuous values of cluster parameters.  
We leave this to a follow-up paper. 

We also emphasize the use of the F1 score as a mechanism for apples-to-apples cluster-finding method comparisons and method combinations.  The F1 score plays a distinct role of enabling a performance comparison between the two methods considered.  The purity and completeness of a method depends on a threshold for positive identification.  Shifts in that threshold for a given method will change the quoted purity and completeness.  We therefore choose a threshold for each method that maximizes the F1 score in that method.  In other words, our cluster-finding comparison compares the best version of the CNN and MF to one another, using the F1 score as a metric for ``best''.  We additionally use the F1 score as a metric that allows us to combine the two cluster-finding methods to further improve performance.  We present a use case of the F1 score as a comparison metric and combination mechanism that can be used as a template for other cluster-finding comparisons and combinations.

Finally, we comment on the complementarity of MF and CNN.  MF inherently relies on an understanding of the expected signal and noise in the filter definition.  The CNN relies on an understanding of the data when generating the training data set, but does not rely on any assumptions in the network architecture nor did it require that the data undergo preprocessing steps such as point source removal.  Another distinguishing aspect between the two is that the MF has an explicit fully analytic formulation as a discriminative model, while the CNN has a quasi-analytic (semi-parametric) formulation via the simulation data used to train the model. A common criticism of machine learning methods, particularly neural networks, is in the lack of interpretability.  While the MF method has physically motivated, or a priori physically denoted, features the method is designed to detect, the CNN picks up on a non-linear combination of features that do not necessarily have obvious physical motivation. Last, we note that the MF method had taken human time to calibrate.  While use of the CNN did not require significant calibration, the human time went into developing the CNN architecture and training the model.

\figureSelFuncBoth

\section{Acknowledgements}

{\it Author Contributions}
\begin{itemize}
    \item \textbf{Lin}: led the design and experiments of the standalone NN and ensemble methods, prepared the cutouts from raw data after appropriate checks, performed the evaluation and diagnostic analysis for MF, NN and ensemble models, contributed to manuscript writing, and kept the work going through challenging phases. 
    \item \textbf{Huang}: optimized and ran the matched filter, producing a catalog for comparison, and contributed to manuscript writing.
    \item \textbf{Avestruz}: contributed to the initial project conception, project management, manuscript writing, and student mentorship, including guidance on analysis direction.
    \item \textbf{Wu}: generated simulations for testing and validation; did initial classification; contributed to manuscript writing.    
    \item \textbf{Trivedi}: prototyped initial architecture and experiments, provided guidance and student mentorship on subsequent deep learning components, including sharing expertise on ML methodology.
    \item \textbf{Caldeira}: contributed to comparisons between results of the two methods and manuscript writing.
    \item \textbf{Nord}: developed the initial project conception, helped assemble the research team, contributed to technical development, project management, manuscript writing, and student mentorship, including guidance on analysis direction.
\end{itemize}

We would like to thank Arya Farahi for discussions that helped improve this manuscript.  We also thank Phil Mansfield for help on manuscript presentation at the end of the project.  CA would like to thank support from the LSA Collegiate Fellows program and the Leinweber Center for Theoretical Physics at the University of Michigan.  This project was supported in part by NSF-AAG awards AST-200994 and AST-2009121. During part of the project, ST was supported by by the National Science Foundation under Grant No. DMS-1439786 while he was in residence at the Institute for Computational and Experimental Research in Mathematics in Providence, RI, during the non-linear algebra program.

We acknowledge the \href{Deep Skies Lab}{https://deepskieslab.com} as a community of multi-domain experts and collaborators who have facilitated an environment of open discussion, idea-generation, and collaboration. 
This community was important for the development of this project.  We would also like to extend our thanks to the SPT collaboration for providing the matched filter implementation.

This manuscript has been authored by Fermi Research Alliance, LLC under Contract No. DE-AC02-07CH11359 with the U.S. Department of Energy, Office of Science, Office of High Energy Physics.
This material is based upon work supported by the National Science Foundation Graduate Research Fellowship under Grant No. DGE 1752814.

\newpage

\newpage









\clearpage

\bibliographystyle{model2-names} 
\bibliography{bib}

\end{document}


\begin{frontmatter}

\title{DeepSZ: Identification of Sunyaev-Zel'dovich Galaxy Clusters using Deep Learning}



\author[Lin et al.]{
Z. ~Lin$^{1}$\orcidicon{0000-0001-8673-6868}\thanks{Contact e-mail: \href{mailto:zhenlin4@illinois.edu}{zhenlin4@illinois.edu}},
N. ~Huang$^{2},$
C.~Avestruz$^{3,4}$\orcidicon{0000-0001-8868-0810},
W. L. K.~Wu$^{5,6}$\orcidicon{0000-0001-5411-6920},
S.~Trivedi$^{7},$
\newauthor
\text{ }J.~Caldeira$^{8},$
B.~Nord$^{5,8,9}$\orcidicon{0000-0001-6706-8972}\\\\
${^1}$\uiuccs\\
${^2}$\berkeley\\
${^3}$\umichigan\\
${^4}$\umlctp\\
${^5}$\kicp\\
${^6}$\slac\\
${^7}$\mit\\
${^8}$\fnal\\
${^9}$\uchicago\\
}

\date{Last updated 2019 August 24}
\pubyear{2019}


\maketitle
\begin{keywords}
cosmic microwave background \sep
galaxy clusters \sep
cosmology \sep 
deep learning \sep
\end{keywords}

\end{frontmatter}

\section{CNN Network Structure}
\label{sec:appendix_1}

\def \CNNStructureSummary{
\footnotesize

\begin{longtable}[]{llllllllllll}
ID  & name                  & type         & ch\_in & dim\_in & ch\_out & dim\_out \\
1   & data                  & data         & 3      & 32x32   & 3       & 32x32    \\
2   & conv1\_1              & Convolution  & 3      & 32x32   & 64      & 32x32    \\
    &                       &              &        &         &         &          \\
3   & bn\_conv1\_1          & BatchNorm    & 64     & 32x32   & 64      & 32x32    \\
    &                       &              &        &         &         &          \\
4   & scale\_conv1\_1       & Scale        & 64     & 32x32   & 64      & 32x32    \\
5   & conv1\_1\_relu        & ReLU         & 64     & 32x32   & 64      & 32x32    \\
6   & conv1\_2              & Convolution  & 64     & 32x32   & 64      & 16x16    \\
    &                       &              &        &         &         &          \\
7   & bn\_conv1\_2          & BatchNorm    & 64     & 16x16   & 64      & 16x16    \\
    &                       &              &        &         &         &          \\
8   & scale\_conv1\_2       & Scale        & 64     & 16x16   & 64      & 16x16    \\
9   & conv1\_2\_relu        & ReLU         & 64     & 16x16   & 64      & 16x16    \\
10  & pool1                 & Pooling      & 64     & 16x16   & 64      & 8x8      \\
11  & res2a\_branch2a       & Convolution  & 64     & 8x8     & 64      & 8x8      \\
    &                       &              &        &         &         &          \\
12  & bn2a\_branch2a        & BatchNorm    & 64     & 8x8     & 64      & 8x8      \\
    &                       &              &        &         &         &          \\
13  & scale2a\_branch2a     & Scale        & 64     & 8x8     & 64      & 8x8      \\
14  & res2a\_branch2a\_relu & ReLU         & 64     & 8x8     & 64      & 8x8      \\
15  & res2a\_branch2b       & Convolution  & 64     & 8x8     & 64      & 8x8      \\
    &                       &              &        &         &         &          \\
16  & bn2a\_branch2b        & BatchNorm    & 64     & 8x8     & 64      & 8x8      \\
    &                       &              &        &         &         &          \\
17  & scale2a\_branch2b     & Scale        & 64     & 8x8     & 64      & 8x8      \\
18  & res2a\_branch2b\_relu & ReLU         & 64     & 8x8     & 64      & 8x8      \\
19  & res2a\_branch2c       & Convolution  & 64     & 8x8     & 256     & 8x8      \\
    &                       &              &        &         &         &          \\
20  & bn2a\_branch2c        & BatchNorm    & 256    & 8x8     & 256     & 8x8      \\
    &                       &              &        &         &         &          \\
21  & scale2a\_branch2c     & Scale        & 256    & 8x8     & 256     & 8x8      \\
22  & res2a\_branch1        & Convolution  & 64     & 8x8     & 256     & 8x8      \\
    &                       &              &        &         &         &          \\
23  & bn2a\_branch1         & BatchNorm    & 256    & 8x8     & 256     & 8x8      \\
    &                       &              &        &         &         &          \\
24  & scale2a\_branch1      & Scale        & 256    & 8x8     & 256     & 8x8      \\
25  & res2a                 & Eltwise      & 256    & 8x8     & 256     & 8x8      \\
26  & res2a\_relu           & ReLU         & 256    & 8x8     & 256     & 8x8      \\
27  & res2b\_branch2a       & Convolution  & 256    & 8x8     & 64      & 8x8      \\
    &                       &              &        &         &         &          \\
28  & bn2b\_branch2a        & BatchNorm    & 64     & 8x8     & 64      & 8x8      \\
    &                       &              &        &         &         &          \\
29  & scale2b\_branch2a     & Scale        & 64     & 8x8     & 64      & 8x8      \\
30  & res2b\_branch2a\_relu & ReLU         & 64     & 8x8     & 64      & 8x8      \\
31  & res2b\_branch2b       & Convolution  & 64     & 8x8     & 64      & 8x8      \\
    &                       &              &        &         &         &          \\
32  & bn2b\_branch2b        & BatchNorm    & 64     & 8x8     & 64      & 8x8      \\
    &                       &              &        &         &         &          \\
33  & scale2b\_branch2b     & Scale        & 64     & 8x8     & 64      & 8x8      \\
34  & res2b\_branch2b\_relu & ReLU         & 64     & 8x8     & 64      & 8x8      \\
35  & res2b\_branch2c       & Convolution  & 64     & 8x8     & 256     & 8x8      \\
    &                       &              &        &         &         &          \\
36  & bn2b\_branch2c        & BatchNorm    & 256    & 8x8     & 256     & 8x8      \\
    &                       &              &        &         &         &          \\
37  & scale2b\_branch2c     & Scale        & 256    & 8x8     & 256     & 8x8      \\
38  & res2b                 & Eltwise      & 256    & 8x8     & 256     & 8x8      \\
39  & res2b\_relu           & ReLU         & 256    & 8x8     & 256     & 8x8      \\
40  & res2c\_branch2a       & Convolution  & 256    & 8x8     & 64      & 8x8      \\
    &                       &              &        &         &         &          \\
41  & bn2c\_branch2a        & BatchNorm    & 64     & 8x8     & 64      & 8x8      \\
    &                       &              &        &         &         &          \\
42  & scale2c\_branch2a     & Scale        & 64     & 8x8     & 64      & 8x8      \\
43  & res2c\_branch2a\_relu & ReLU         & 64     & 8x8     & 64      & 8x8      \\
44  & res2c\_branch2b       & Convolution  & 64     & 8x8     & 64      & 8x8      \\
    &                       &              &        &         &         &          \\
45  & bn2c\_branch2b        & BatchNorm    & 64     & 8x8     & 64      & 8x8      \\
    &                       &              &        &         &         &          \\
46  & scale2c\_branch2b     & Scale        & 64     & 8x8     & 64      & 8x8      \\
47  & res2c\_branch2b\_relu & ReLU         & 64     & 8x8     & 64      & 8x8      \\
48  & res2c\_branch2c       & Convolution  & 64     & 8x8     & 256     & 8x8      \\
    &                       &              &        &         &         &          \\
49  & bn2c\_branch2c        & BatchNorm    & 256    & 8x8     & 256     & 8x8      \\
    &                       &              &        &         &         &          \\
50  & scale2c\_branch2c     & Scale        & 256    & 8x8     & 256     & 8x8      \\
51  & res2c                 & Eltwise      & 256    & 8x8     & 256     & 8x8      \\
52  & res2c\_relu           & ReLU         & 256    & 8x8     & 256     & 8x8      \\
53  & res3a\_branch2a       & Convolution  & 256    & 8x8     & 128     & 4x4      \\
    &                       &              &        &         &         &          \\
54  & bn3a\_branch2a        & BatchNorm    & 128    & 4x4     & 128     & 4x4      \\
    &                       &              &        &         &         &          \\
55  & scale3a\_branch2a     & Scale        & 128    & 4x4     & 128     & 4x4      \\
56  & res3a\_branch2a\_relu & ReLU         & 128    & 4x4     & 128     & 4x4      \\
57  & res3a\_branch2b       & Convolution  & 128    & 4x4     & 128     & 4x4      \\
    &                       &              &        &         &         &          \\
58  & bn3a\_branch2b        & BatchNorm    & 128    & 4x4     & 128     & 4x4      \\
    &                       &              &        &         &         &          \\
59  & scale3a\_branch2b     & Scale        & 128    & 4x4     & 128     & 4x4      \\
60  & res3a\_branch2b\_relu & ReLU         & 128    & 4x4     & 128     & 4x4      \\
61  & res3a\_branch2c       & Convolution  & 128    & 4x4     & 512     & 4x4      \\
    &                       &              &        &         &         &          \\
62  & bn3a\_branch2c        & BatchNorm    & 512    & 4x4     & 512     & 4x4      \\
    &                       &              &        &         &         &          \\
63  & scale3a\_branch2c     & Scale        & 512    & 4x4     & 512     & 4x4      \\
64  & res3a\_branch1        & Convolution  & 256    & 8x8     & 512     & 4x4      \\
    &                       &              &        &         &         &          \\
65  & bn3a\_branch1         & BatchNorm    & 512    & 4x4     & 512     & 4x4      \\
    &                       &              &        &         &         &          \\
66  & scale3a\_branch1      & Scale        & 512    & 4x4     & 512     & 4x4      \\
67  & res3a                 & Eltwise      & 512    & 4x4     & 512     & 4x4      \\
68  & res3a\_relu           & ReLU         & 512    & 4x4     & 512     & 4x4      \\
69  & res3b\_branch2a       & Convolution  & 512    & 4x4     & 128     & 4x4      \\
    &                       &              &        &         &         &          \\
70  & bn3b\_branch2a        & BatchNorm    & 128    & 4x4     & 128     & 4x4      \\
    &                       &              &        &         &         &          \\
71  & scale3b\_branch2a     & Scale        & 128    & 4x4     & 128     & 4x4      \\
72  & res3b\_branch2a\_relu & ReLU         & 128    & 4x4     & 128     & 4x4      \\
73  & res3b\_branch2b       & Convolution  & 128    & 4x4     & 128     & 4x4      \\
    &                       &              &        &         &         &          \\
74  & bn3b\_branch2b        & BatchNorm    & 128    & 4x4     & 128     & 4x4      \\
    &                       &              &        &         &         &          \\
75  & scale3b\_branch2b     & Scale        & 128    & 4x4     & 128     & 4x4      \\
76  & res3b\_branch2b\_relu & ReLU         & 128    & 4x4     & 128     & 4x4      \\
77  & res3b\_branch2c       & Convolution  & 128    & 4x4     & 512     & 4x4      \\
    &                       &              &        &         &         &          \\
78  & bn3b\_branch2c        & BatchNorm    & 512    & 4x4     & 512     & 4x4      \\
    &                       &              &        &         &         &          \\
79  & scale3b\_branch2c     & Scale        & 512    & 4x4     & 512     & 4x4      \\
80  & res3b                 & Eltwise      & 512    & 4x4     & 512     & 4x4      \\
81  & res3b\_relu           & ReLU         & 512    & 4x4     & 512     & 4x4      \\
82  & res3c\_branch2a       & Convolution  & 512    & 4x4     & 128     & 4x4      \\
    &                       &              &        &         &         &          \\
83  & bn3c\_branch2a        & BatchNorm    & 128    & 4x4     & 128     & 4x4      \\
    &                       &              &        &         &         &          \\
84  & scale3c\_branch2a     & Scale        & 128    & 4x4     & 128     & 4x4      \\
85  & res3c\_branch2a\_relu & ReLU         & 128    & 4x4     & 128     & 4x4      \\
86  & res3c\_branch2b       & Convolution  & 128    & 4x4     & 128     & 4x4      \\
    &                       &              &        &         &         &          \\
87  & bn3c\_branch2b        & BatchNorm    & 128    & 4x4     & 128     & 4x4      \\
    &                       &              &        &         &         &          \\
88  & scale3c\_branch2b     & Scale        & 128    & 4x4     & 128     & 4x4      \\
89  & res3c\_branch2b\_relu & ReLU         & 128    & 4x4     & 128     & 4x4      \\
90  & res3c\_branch2c       & Convolution  & 128    & 4x4     & 512     & 4x4      \\
    &                       &              &        &         &         &          \\
91  & bn3c\_branch2c        & BatchNorm    & 512    & 4x4     & 512     & 4x4      \\
    &                       &              &        &         &         &          \\
92  & scale3c\_branch2c     & Scale        & 512    & 4x4     & 512     & 4x4      \\
93  & res3c                 & Eltwise      & 512    & 4x4     & 512     & 4x4      \\
94  & res3c\_relu           & ReLU         & 512    & 4x4     & 512     & 4x4      \\
95  & res3d\_branch2a       & Convolution  & 512    & 4x4     & 128     & 4x4      \\
    &                       &              &        &         &         &          \\
96  & bn3d\_branch2a        & BatchNorm    & 128    & 4x4     & 128     & 4x4      \\
    &                       &              &        &         &         &          \\
97  & scale3d\_branch2a     & Scale        & 128    & 4x4     & 128     & 4x4      \\
98  & res3d\_branch2a\_relu & ReLU         & 128    & 4x4     & 128     & 4x4      \\
99  & res3d\_branch2b       & Convolution  & 128    & 4x4     & 128     & 4x4      \\
    &                       &              &        &         &         &          \\
100 & bn3d\_branch2b        & BatchNorm    & 128    & 4x4     & 128     & 4x4      \\
    &                       &              &        &         &         &          \\
101 & scale3d\_branch2b     & Scale        & 128    & 4x4     & 128     & 4x4      \\
102 & res3d\_branch2b\_relu & ReLU         & 128    & 4x4     & 128     & 4x4      \\
103 & res3d\_branch2c       & Convolution  & 128    & 4x4     & 512     & 4x4      \\
    &                       &              &        &         &         &          \\
104 & bn3d\_branch2c        & BatchNorm    & 512    & 4x4     & 512     & 4x4      \\
    &                       &              &        &         &         &          \\
105 & scale3d\_branch2c     & Scale        & 512    & 4x4     & 512     & 4x4      \\
106 & res3d                 & Eltwise      & 512    & 4x4     & 512     & 4x4      \\
107 & res3d\_relu           & ReLU         & 512    & 4x4     & 512     & 4x4      \\
108 & res4a\_branch2a       & Convolution  & 512    & 4x4     & 256     & 2x2      \\
    &                       &              &        &         &         &          \\
109 & bn4a\_branch2a        & BatchNorm    & 256    & 2x2     & 256     & 2x2      \\
    &                       &              &        &         &         &          \\
110 & scale4a\_branch2a     & Scale        & 256    & 2x2     & 256     & 2x2      \\
111 & res4a\_branch2a\_relu & ReLU         & 256    & 2x2     & 256     & 2x2      \\
112 & res4a\_branch2b       & Convolution  & 256    & 2x2     & 256     & 2x2      \\
    &                       &              &        &         &         &          \\
113 & bn4a\_branch2b        & BatchNorm    & 256    & 2x2     & 256     & 2x2      \\
    &                       &              &        &         &         &          \\
114 & scale4a\_branch2b     & Scale        & 256    & 2x2     & 256     & 2x2      \\
115 & res4a\_branch2b\_relu & ReLU         & 256    & 2x2     & 256     & 2x2      \\
116 & res4a\_branch2c       & Convolution  & 256    & 2x2     & 1024    & 2x2      \\
    &                       &              &        &         &         &          \\
117 & bn4a\_branch2c        & BatchNorm    & 1024   & 2x2     & 1024    & 2x2      \\
    &                       &              &        &         &         &          \\
118 & scale4a\_branch2c     & Scale        & 1024   & 2x2     & 1024    & 2x2      \\
119 & res4a\_branch1        & Convolution  & 512    & 4x4     & 1024    & 2x2      \\
    &                       &              &        &         &         &          \\
120 & bn4a\_branch1         & BatchNorm    & 1024   & 2x2     & 1024    & 2x2      \\
    &                       &              &        &         &         &          \\
121 & scale4a\_branch1      & Scale        & 1024   & 2x2     & 1024    & 2x2      \\
122 & res4a                 & Eltwise      & 1024   & 2x2     & 1024    & 2x2      \\
123 & res4a\_relu           & ReLU         & 1024   & 2x2     & 1024    & 2x2      \\
124 & res4b\_branch2a       & Convolution  & 1024   & 2x2     & 256     & 2x2      \\
    &                       &              &        &         &         &          \\
125 & bn4b\_branch2a        & BatchNorm    & 256    & 2x2     & 256     & 2x2      \\
    &                       &              &        &         &         &          \\
126 & scale4b\_branch2a     & Scale        & 256    & 2x2     & 256     & 2x2      \\
127 & res4b\_branch2a\_relu & ReLU         & 256    & 2x2     & 256     & 2x2      \\
128 & res4b\_branch2b       & Convolution  & 256    & 2x2     & 256     & 2x2      \\
    &                       &              &        &         &         &          \\
129 & bn4b\_branch2b        & BatchNorm    & 256    & 2x2     & 256     & 2x2      \\
    &                       &              &        &         &         &          \\
130 & scale4b\_branch2b     & Scale        & 256    & 2x2     & 256     & 2x2      \\
131 & res4b\_branch2b\_relu & ReLU         & 256    & 2x2     & 256     & 2x2      \\
132 & res4b\_branch2c       & Convolution  & 256    & 2x2     & 1024    & 2x2      \\
    &                       &              &        &         &         &          \\
133 & bn4b\_branch2c        & BatchNorm    & 1024   & 2x2     & 1024    & 2x2      \\
    &                       &              &        &         &         &          \\
134 & scale4b\_branch2c     & Scale        & 1024   & 2x2     & 1024    & 2x2      \\
135 & res4b                 & Eltwise      & 1024   & 2x2     & 1024    & 2x2      \\
136 & res4b\_relu           & ReLU         & 1024   & 2x2     & 1024    & 2x2      \\
137 & res4c\_branch2a       & Convolution  & 1024   & 2x2     & 256     & 2x2      \\
    &                       &              &        &         &         &          \\
138 & bn4c\_branch2a        & BatchNorm    & 256    & 2x2     & 256     & 2x2      \\
    &                       &              &        &         &         &          \\
139 & scale4c\_branch2a     & Scale        & 256    & 2x2     & 256     & 2x2      \\
140 & res4c\_branch2a\_relu & ReLU         & 256    & 2x2     & 256     & 2x2      \\
141 & res4c\_branch2b       & Convolution  & 256    & 2x2     & 256     & 2x2      \\
    &                       &              &        &         &         &          \\
142 & bn4c\_branch2b        & BatchNorm    & 256    & 2x2     & 256     & 2x2      \\
    &                       &              &        &         &         &          \\
143 & scale4c\_branch2b     & Scale        & 256    & 2x2     & 256     & 2x2      \\
144 & res4c\_branch2b\_relu & ReLU         & 256    & 2x2     & 256     & 2x2      \\
145 & res4c\_branch2c       & Convolution  & 256    & 2x2     & 1024    & 2x2      \\
    &                       &              &        &         &         &          \\
146 & bn4c\_branch2c        & BatchNorm    & 1024   & 2x2     & 1024    & 2x2      \\
    &                       &              &        &         &         &          \\
147 & scale4c\_branch2c     & Scale        & 1024   & 2x2     & 1024    & 2x2      \\
148 & res4c                 & Eltwise      & 1024   & 2x2     & 1024    & 2x2      \\
149 & res4c\_relu           & ReLU         & 1024   & 2x2     & 1024    & 2x2      \\
150 & res4d\_branch2a       & Convolution  & 1024   & 2x2     & 256     & 2x2      \\
    &                       &              &        &         &         &          \\
151 & bn4d\_branch2a        & BatchNorm    & 256    & 2x2     & 256     & 2x2      \\
    &                       &              &        &         &         &          \\
152 & scale4d\_branch2a     & Scale        & 256    & 2x2     & 256     & 2x2      \\
153 & res4d\_branch2a\_relu & ReLU         & 256    & 2x2     & 256     & 2x2      \\
154 & res4d\_branch2b       & Convolution  & 256    & 2x2     & 256     & 2x2      \\
    &                       &              &        &         &         &          \\
155 & bn4d\_branch2b        & BatchNorm    & 256    & 2x2     & 256     & 2x2      \\
    &                       &              &        &         &         &          \\
156 & scale4d\_branch2b     & Scale        & 256    & 2x2     & 256     & 2x2      \\
157 & res4d\_branch2b\_relu & ReLU         & 256    & 2x2     & 256     & 2x2      \\
158 & res4d\_branch2c       & Convolution  & 256    & 2x2     & 1024    & 2x2      \\
    &                       &              &        &         &         &          \\
159 & bn4d\_branch2c        & BatchNorm    & 1024   & 2x2     & 1024    & 2x2      \\
    &                       &              &        &         &         &          \\
160 & scale4d\_branch2c     & Scale        & 1024   & 2x2     & 1024    & 2x2      \\
161 & res4d                 & Eltwise      & 1024   & 2x2     & 1024    & 2x2      \\
162 & res4d\_relu           & ReLU         & 1024   & 2x2     & 1024    & 2x2      \\
163 & res4e\_branch2a       & Convolution  & 1024   & 2x2     & 256     & 2x2      \\
    &                       &              &        &         &         &          \\
164 & bn4e\_branch2a        & BatchNorm    & 256    & 2x2     & 256     & 2x2      \\
    &                       &              &        &         &         &          \\
165 & scale4e\_branch2a     & Scale        & 256    & 2x2     & 256     & 2x2      \\
166 & res4e\_branch2a\_relu & ReLU         & 256    & 2x2     & 256     & 2x2      \\
167 & res4e\_branch2b       & Convolution  & 256    & 2x2     & 256     & 2x2      \\
    &                       &              &        &         &         &          \\
168 & bn4e\_branch2b        & BatchNorm    & 256    & 2x2     & 256     & 2x2      \\
    &                       &              &        &         &         &          \\
169 & scale4e\_branch2b     & Scale        & 256    & 2x2     & 256     & 2x2      \\
170 & res4e\_branch2b\_relu & ReLU         & 256    & 2x2     & 256     & 2x2      \\
171 & res4e\_branch2c       & Convolution  & 256    & 2x2     & 1024    & 2x2      \\
    &                       &              &        &         &         &          \\
172 & bn4e\_branch2c        & BatchNorm    & 1024   & 2x2     & 1024    & 2x2      \\
    &                       &              &        &         &         &          \\
173 & scale4e\_branch2c     & Scale        & 1024   & 2x2     & 1024    & 2x2      \\
174 & res4e                 & Eltwise      & 1024   & 2x2     & 1024    & 2x2      \\
175 & res4e\_relu           & ReLU         & 1024   & 2x2     & 1024    & 2x2      \\
176 & res4f\_branch2a       & Convolution  & 1024   & 2x2     & 256     & 2x2      \\
    &                       &              &        &         &         &          \\
177 & bn4f\_branch2a        & BatchNorm    & 256    & 2x2     & 256     & 2x2      \\
    &                       &              &        &         &         &          \\
178 & scale4f\_branch2a     & Scale        & 256    & 2x2     & 256     & 2x2      \\
179 & res4f\_branch2a\_relu & ReLU         & 256    & 2x2     & 256     & 2x2      \\
180 & res4f\_branch2b       & Convolution  & 256    & 2x2     & 256     & 2x2      \\
    &                       &              &        &         &         &          \\
181 & bn4f\_branch2b        & BatchNorm    & 256    & 2x2     & 256     & 2x2      \\
    &                       &              &        &         &         &          \\
182 & scale4f\_branch2b     & Scale        & 256    & 2x2     & 256     & 2x2      \\
183 & res4f\_branch2b\_relu & ReLU         & 256    & 2x2     & 256     & 2x2      \\
184 & res4f\_branch2c       & Convolution  & 256    & 2x2     & 1024    & 2x2      \\
    &                       &              &        &         &         &          \\
185 & bn4f\_branch2c        & BatchNorm    & 1024   & 2x2     & 1024    & 2x2      \\
    &                       &              &        &         &         &          \\
186 & scale4f\_branch2c     & Scale        & 1024   & 2x2     & 1024    & 2x2      \\
187 & res4f                 & Eltwise      & 1024   & 2x2     & 1024    & 2x2      \\
188 & res4f\_relu           & ReLU         & 1024   & 2x2     & 1024    & 2x2      \\
189 & bn4f                  & BatchNorm    & 1024   & 2x2     & 1024    & 2x2      \\
    &                       &              &        &         &         &          \\
190 & scale4f               & Scale        & 1024   & 2x2     & 1024    & 2x2      \\
191 & fc1                   & InnerProduct & 1024   & 2x2     & 256     & 1x1      \\
    &                       &              &        &         &         &          \\
192 & fc1\_relu             & ReLU         & 256    & 1x1     & 256     & 1x1      \\
193 & fc2                   & InnerProduct & 256    & 1x1     & 256     & 1x1      \\
    &                       &              &        &         &         &          \\
194 & fc2\_relu             & ReLU         & 256    & 1x1     & 256     & 1x1      \\
195 & prob\_fc              & InnerProduct & 256    & 1x1     & 2       & 1x1      \\
    &                       &              &        &         &         &          \\
196 & prob\_relu            & ReLU         & 2      & 1x1     & 2       & 1x1      \\
197 & prob                  & Softmax      & 2      & 1x1     & 2       & 1x1      \\

\end{longtable}

}
We base our model structure on ResNet-50 \textcolor{red}{Ref}. 
Because our image dimension is much smaller compared to ResNet-50's task, we made changes including decreasing the convolution kernel sizes in the stem and reducing the number of residual blocks. A full summary of the neural net can be found below.
\CNNStructureSummary